\documentclass[aps,preprint,floatfix,nofootinbib,showpacs]{revtex4-1}
\pdfoutput=1
\usepackage{graphicx,color}
\usepackage{hyperref}
\usepackage{amsmath}
\usepackage{amsfonts}
\usepackage{amssymb}
\usepackage{setspace}

\begin{document}
\title{Detecting multimuon-jets from the Higgs exotic decays in the 
Higgs portal framework}

\renewcommand{\thefootnote}{\arabic{footnote}}

\author{
Jung Chang$^1$, Kingman Cheung$^{1,2,3}$, Shih-Chieh Hsu$^{3,4}$, 
and Chih-Ting Lu$^3$
}
\affiliation{
$^1$ Physics Division, National Center for Theoretical Sciences,
Hsinchu 300, Taiwan \\
$^2$ Division of Quantum Phases and Devices, School of Physics, 
Konkuk University, Seoul 143-701, Republic of Korea \\
$^3$ Department of Physics, National Tsing Hua University,
Hsinchu 300, Taiwan \\
$^4$ Department of Physics, University of Washington at Seattle, Seattle,
WA 98195, U.S.A
}
\date{\today}

\begin{abstract}
A muon-jet ($\mu$-jet) is a very special feature that consists of a
cluster of collimated muons from the decay of a fast moving light
particle of mass about $\mathcal{O}$(1 GeV). We will use this 
feature to search for
very light particles from rare decays of the Higgs boson.
For such a small angular separation of muons which might 
come from a long-lived particle, both ATLAS and CMS could have 
the displaced-vertexing-reconstruction capability.
We use two simple models of the Higgs-portal type to explore the
possibilities of event topologies with two $2\mu$-jets, one $2\mu$-jet
$\&$ one $4\mu$-jet, and two $4\mu$-jets in the final state at LHC-14. We
also summarize existing constraints on these models.
\end{abstract}

\maketitle

\section{Introduction}
A lepton-jet is an object that consists of a group of boosted and 
collimated leptons. It is a unique signature for the leptonic 
decay of a fast moving light particle in the mass range below about
1 GeV \cite{hidden}, where it was discussed in the context of light scalar bosons and gauge bosons from the dark sector.

In these dark-sector models, 
the Higgs boson can be connected the dark sector via Higgs-portal type 
interaction: $(\Phi^\dagger \Phi) (S^\dagger S)$, where $\Phi$ is the 
standard model (SM) Higgs field and $S$ is the scalar field in the dark sector.  When both the Higgs field and $S$ develop vacuum expectation 
values,
the $\Phi$ and $S$ mix to form mass eigenstates, and the Higgs boson 
can decay into a pair of the scalar bosons if kinematically allowed.
In some models, the dark sector can also be connected with the SM particles
via $Z-Z'$ mixing. In either scenarios, when the dark scalar bosons or
gauge bosons are very light, say below 1 GeV, they will decay
into the heaviest SM particles if kinematically allowed.  For example,
a 500 MeV scalar boson decays, via the mixing with the SM Higgs boson, 
can decay into a pair of muons, pions, electrons, or photons. The 
dominant modes would be pions and muons. In some other scenarios 
when there is a broken $U(1)$ global symmetry in the dark sector, 
the scalar boson can also decay into a pair of Goldstone bosons
\cite{gb}.
A UV complete model, which can have one light
pseudoscalar resonance ($a_{1}$) decaying into a pair of muons, is
the next-to-minimal supersymmetric standard model (NMSSM) \cite{NMSSM}. 
Both the dark-sector models and the NMSSM have been frequently explored
in the LHC experiments: ATLAS \cite{lj-ATLAS} and CMS \cite{lj-CMS}.

In this work,
we focus on the search for muon-jets from the decay of very light particles
so as to identify the existence of dark sectors that are connected to the
SM via the Higgs-portal.  For simplicity we only consider the 
dark-sector models that contain either a real SM-singlet scalar 
field $X$ or two real SM-singlet scalar fields $X_{1}$ and $X_{2}$,
without imposing any extra symmetries. The dominant decay modes of the
scalar boson of $\mathcal{O}$(1 GeV) would then be pions and muons.  It is 
the muons 
in the decay products of a fast-moving light scalar boson
that form a muon-jet, for which we are searching for 
in experiments as a signature of the existence of very light scalar bosons.
Such a light scalar boson, which originates from the mixing with the SM
Higgs boson, can appear in the decay of the Higgs boson. 
Since the constraints on the mixing for such a light 
scalar boson with the SM Higgs boson are very strong, which we will
show, the very light scalar boson might be a long-lived neutral
particle and so we might observe a displaced decay vertex 
in detectors.
We are therefore interested in rare decays of the Higgs boson 
into a pair of very light scalar bosons of mass about 
$\mathcal{O}$(1 GeV), each of
which in turns directly decays into a pair of
collimated muons or in a cascade decay into 
another pair of light scalar bosons, then each of them decays into a pair of 
collimated muons.
Let us denote a muon-jet with $n$ collimated muons in it by 
$n\mu$-jet, e.g., a $2\mu$-jet is a muon-jet with 2 muons and 
a $4\mu$-jet is a muon-jet with 4 muons. 
Thus, the final states can consist of three different types
of combinations : \\
(1) two $2\mu$-jets back-to-back in the transverse plane; \\
(2) one $2\mu$-jet on one side and one $4\mu$-jet on the other side; \\
(3) two $4\mu$-jets on opposite sides of the transverse plane.

The model can be made further complicated by invoking additional 
light scalar bosons or gauge bosons in the dark
sector such that the decay chain can involve more dark sector
particles. At the end, each lepton-jet can consist of more than four
leptons, like 6, 8, or more. These lepton-jets would be very 
interesting objects to search for in experiments because 
they are clear signals of new physics. The experimental resolution
to tell the number of leptons in a lepton-jet becomes an important
issue. 
Here we only consider two simple cases of two muons inside a muon-jet
and four muons inside a ``fat'' muon-jet. We also
compare these two cases to distinguish between 
whether the final state muon-jet is produced 
from direct decay of a light scalar boson or in a cascade decay.

The main goal of this work is to investigate the capability of the LHC 
detectors (especially the ATLAS because of its size) of observing muon-jets
in search of rare decays of the Higgs bosons into a pair of very light
bosons, which may decay directly into muon-jets or indirectly 
via subsequent decays into other lighter bosons.  We found that if the
light Higgs boson heavier than about 0.3 GeV, the ATLAS detector 
has a good chance of seeing that.

We would like to reminder the readers that the model considered in this
work is only a toy-model of the Higgs-portal type models.
The other popular models such as dark-$Z$ and dark-photon fall in the same 
category of models in the present context. Nevertheless, 
the search for dark-photon or dark-$Z$ also includes the electron-jets
and pion-jets. The choice depends on the branching ratios and also the
capability of the detector. In the present work, we simply focus
on the best capability of observing muon-jets using the tracker and muon
spectrometer at the ATLAS or CMS detector.

The organization is as follows.  
We describe two simple Higgs-portal models in the next
section, and in Sec. III the existing constraints on these two models. These
constraints are quite generic on many models of this kind.  
We consider some kinematical distributions at the LHC
for some benchmark points in Sec. IV and highlight the potential 
search at LHC-14 with 300 $fb^{-1}$ in Sec. V. 
Finally, we conclude in Sec. VI.

\section{Higgs-portal Models}
\subsection{Higgs-portal Model-1 : Only one light scalar $ h_{s} $}

Here we employ a Higgs portal model in which the SM Higgs field $\Phi$ 
can mix with a 
real scalar field $X$. This additional scalar field does not 
have any SM interactions. We also impose a $ Z_{2} $ symmetry which 
$\Phi$ is $ Z_{2}-even $ and $X$ is $ Z_{2}-odd $ before both the Higgs 
field and $X$ develop vacuum expectation values.
The renormalizable Lagrangian density for this model is given by
\begin{eqnarray}
{\cal L} &=& \frac{1}{2}\partial_{\mu}X\partial^{\mu}X
+\frac{1}{2}\mu^{2}_{X}X^{2}-\frac{1}{4}\lambda_{X}X^{4}
-\frac{1}{2}\lambda_{\Phi X}(\Phi^\dagger \Phi)X^{2} \nonumber \\
&+& {\cal L_{SM}} \;, 
\label{L'}
\end{eqnarray}
where the Higgs sector in the $ {\cal L_{SM}} $ is
\begin{eqnarray}
{\cal L_{SM}} &\supset& (D_{\mu}\Phi)^{\dagger}(D^{\mu}\Phi)
 +\mu^{2}(\Phi^{\dagger}\Phi)-\lambda (\Phi^{\dagger}\Phi)^{2} \;.
\end{eqnarray}
After the electroweak symmetry breaking (EWSB), the SM Higgs 
doublet field $\Phi$ is expanded around its vacuum-expectation value:
\begin{equation}
\Phi (x) = \frac{1}{\sqrt{2}} \left( \begin{array}{c}
           0 \\
           \langle \phi \rangle + \phi(x) \end{array} \right ) \;
\end{equation}
where $\langle\phi\rangle\approx 246$ GeV. The $X$ is also expanded 
around its vacuum-expectation value $ \langle\chi\rangle $:
\begin{equation}
X(x) = \langle\chi\rangle + \chi (x) \;
\end{equation}
Two tadpole conditions can be written down using 
$ \partial V/ \partial\phi = 0 $ and $ \partial V/ \partial\chi = 0 $,
where $V$ is the scalar potential part of Eq.~(\ref{L'}):
\begin{equation}
\langle\phi\rangle^{2} = \frac{4\lambda_{X}\mu^{2}-2\lambda_{\Phi X}
\mu_{X}^{2}}{4\lambda\lambda_{X}-\lambda_{\Phi X}^{2}} \;,
\end{equation}
\begin{equation}
\langle\chi\rangle^{2} = \frac{4\lambda\mu_{X}^{2}-2\lambda_{\Phi X}
\mu^{2}}{4\lambda\lambda_{X}-\lambda_{\Phi X}^{2}} \;
\end{equation}
Taking the decoupling limit $ \lambda_{\Phi X}\rightarrow 0 $ 
from the above equations, we recover the SM condition of 
$ \langle\phi\rangle^{2}=\mu^{2}/ \lambda $ as well as 
$ \langle\chi\rangle^{2}=\mu_{X}^{2}/ \lambda_{X} $.

It is easy to see that the Higgs boson field $\phi$ will mix with the
new scalar field $\chi$ to form mass eigenstates denoted by $h$ and $h_{s}$,
respectively. The mass terms for the Higgs boson and the new scalar 
boson are
\begin{equation}
{\cal L}_m =  - \frac{1}{2} \left( \phi \; \chi \right )\,
 \left( \begin{array}{cc} 
         2\lambda\langle\phi\rangle^2 & \lambda_{\Phi X}\langle\phi\rangle\langle\chi\rangle  \\
         \lambda_{\Phi X}\langle\phi\rangle\langle\chi\rangle & 2\lambda_{X}\langle\chi\rangle^2 \end{array} \right )\,
  \left( \begin{array}{c} 
           \phi \\
           \chi \end{array} \right ) \;,
\end{equation}
We can rotate $(\phi\; \chi)^T \longrightarrow (h \; h_{s})^T$ 
through an angle $\theta$
\begin{equation}
 \left( \begin{array}{c}
                h \\ 
                h_{s}  \end{array} \right ) 
= 
 \left( \begin{array}{cc} 
            \cos\theta & \sin \theta \\
            - \sin \theta & \cos\theta  \end{array} \right )\,
 \left( \begin{array}{c}
                \phi \\ 
                \chi  \end{array} \right ) 
\end{equation}
Thus, the masses of the Higgs boson $h$ and the scalar boson $h_{s}$,
the mixing angle $\theta$, and the interaction governing $h \to h_{s} h_{s}$
are given by, in terms of the parameters in Eq.~(\ref{L'}),
\begin{eqnarray}
m_h^2 &=& 2\lambda\langle\phi\rangle^{2}\cos^2\theta +2\lambda_{X}\langle\chi\rangle^{2} \sin^2\theta + \lambda_{\Phi X}\langle\phi\rangle\langle\chi\rangle\sin 2\theta 
 \nonumber \\
m_{h_{s}}^2 &=& 2\lambda_{X}\langle\chi\rangle^{2}\cos^2\theta +2\lambda\langle\phi\rangle^{2}\sin^2\theta - \lambda_{\Phi X}\langle\phi\rangle\langle\chi\rangle\sin 2\theta 
 \nonumber \\
{\cal L}_{hh_{s}h_{s}} &=& -\frac{1}{2}[6\lambda_{X}\langle\chi\rangle\cos ^{2}\theta \sin\theta +6\lambda\langle\phi\rangle\cos\theta\sin^{2}\theta + \lambda_{\Phi X}\langle\phi\rangle (\cos ^{3}\theta-2\cos\theta \sin^{2}\theta)\nonumber\\ 
&&+ \lambda_{\Phi X}\langle\chi\rangle (\sin ^{3}\theta-2\cos^{2}\theta \sin\theta) ]hh_{s}h_{s} \nonumber \\
\tan 2\theta &=& \frac{\lambda_{\Phi X}\langle\phi\rangle\langle\chi\rangle}{\lambda\langle\phi\rangle^{2} - \lambda_{X}\langle\chi\rangle^{2}}
\nonumber 
\end{eqnarray}
In the next section, where we describe the constraints on the model, 
the angle $\theta$ has to be very small.  In the small $\theta$ limit, the
above relations are reduced to
\begin{eqnarray}
m_h^2 & \simeq & 2\lambda\langle\phi\rangle^{2} = (125\;{\rm GeV} )^2 \nonumber \\
m_{h_{s}}^2 & \simeq & 2\lambda_{X}\langle\chi\rangle^{2} \nonumber \\
{\cal L}_{hh_{s}h_{s}} &=& -\frac{1}{2}\lambda_{\Phi X}\langle\phi\rangle h h_{s} h_{s} \nonumber \\
\theta & \simeq & \frac{\lambda_{\Phi X}\langle\phi\rangle\langle\chi\rangle}{m_{h}^{2} - m_{h_{s}}^2} \nonumber \;,
\end{eqnarray}

The scalar boson $h_{s}$ can decay into SM particles via the mixing with
the Higgs boson. Thus, the decay widths for $h_{s} \to \ell^+ \ell^-$ and 
$h_{s} \to \pi \pi$ are given by \cite{Gunion:1989we}
\begin{eqnarray}
\Gamma ( h_{s} \to \ell^+ \ell^-) &=& sin^2\theta \, 
  \frac{m_\ell^2 m_{h_{s}}}{ 8 \pi \langle \phi \rangle^2 } \left( 1 - 
  \frac{4 m_\ell^2}{m_{h_{s}}^2 } \right )^{3/2}\;, \\
\Gamma ( h_{s} \to \pi \pi) &=& sin^2\theta \, 
  \frac{m_{h_{s}}^3}{ 216 \pi \langle \phi \rangle^2 } \,
 \left( 1 -   \frac{4 m_\pi^2}{m_{h_{s}}^2 } \right )^{1/2} \,
 \left( 1 +  \frac{11 m_\pi^2}{2 m_{h_{s}}^2 } \right )^2 \;, \\
\Gamma_{h_{s}} &=& \frac{1}{\tau_{h_{s}}} = \sum_{\ell = e, \mu} 
  \Gamma(h_{s} \to \ell^+ \ell^- )
          +  \sum_{\pi\pi= \pi^+\pi^-,\pi^0\pi^0} \Gamma(h_{s} \to \pi \pi )  \;, 
\end{eqnarray}
where we have restricted $m_{h_{s}} \alt 1$ GeV. 
\footnote
{ Even though the major decay mode of $m_{h_s}=0.3-1$ GeV
    is $\pi\pi$ mode,we still focus on the analysis of
    $\mu^{+}\mu^{-}$ mode. Since the resolution of muons are better
    than pions and the analysis of $\pi\pi$ mode has been researched
    in Ref.\cite{gb,pi-jet}.}
Here $\pi\pi$ includes both $\pi^+\pi^-$ and $\pi^0\pi^0$,
 and $\Gamma (h_{s}\to\pi^{+}\pi^{-})=2\Gamma
(h_{s}\to\pi^{0}\pi^{0})$. Since the tree-level estimate of $\Gamma
(h_{s} \to \pi \pi)$ is not adequate
 when $ m_{h_{s}} $ is not far from
the pion threshold, where the strong final-state interaction becomes
important \cite{Donoghue:1990xh,Bezrukov:2009yw}, so we follow
Ref.~\cite{Donoghue:1990xh,Bezrukov:2009yw} for numerical estimates of
$\Gamma (h_{s} \to \pi \pi)$. We show the branching ratios of the
scalar boson $h_{s}$ for the two most dominant modes $\mu^+\mu^-$ and
$\pi\pi$ in Table~\ref{tab:BR-hs} for $m_{h_s}=0.3-1$ GeV.

\begin{table}[h!]
\caption{\small  \label{tab:BR-hs}
The branching ratio for the most two dominant decay modes of the
scalar boson $h_{s}$ for $m_{h_{s}} =0.3 - 1$ GeV.
Here $\pi\pi$ includes $\pi^+ \pi^-$ and $\pi^0 \pi^0$.
}
\vspace{1.0mm}
\begin{ruledtabular}
 \begin{tabular}{ l c c c c c c c c }
$ m_{h_{s}}$ (GeV) & 0.3 & 0.4 & 0.5 & 0.6 & 0.7 & 0.8 & 0.9 & 1.0 \\ \hline
$ B(\mu^+\mu^-) $ & $ 40\% $ & $ 12.5\% $ & $ 10\% $ & $ 8\% $ & $ 6\% $ & $ 4.5\% $ & $ 1.5\% $ & $ 0.4\% $ \\ 
$ B(\pi\pi) $ & $ 60\% $ & $ 87.5\% $ & $ 90\% $ & $ 92\% $ & $ 94\% $ & $ 95.5\% $ & $ 98.5\% $ & $ 99.6\% $ \\ \hline
\end{tabular}
\end{ruledtabular}
\end{table}

\subsection{Higgs-portal model-2 : Two light scalars : $ h_{D_1} $, $ h_{D_2} $}

We can extend our Higgs-portal model-1 to include two real scalar 
fields $X_1$ and $X_2$, which 
can mix with the SM Higgs field but do not have any SM interactions.
We also impose a $ Z_{2} $ symmetry which $\Phi$ is $ Z_{2}-even $ and 
both $X_1$, $X_2$ are $ Z_{2}-odd $ before these Higgs field, 
$X_1$ and $X_2$ develop vacuum expectation values.
The renormalizable Lagrangian density for this model is given by
\begin{eqnarray}
{\cal L} &=& \frac{1}{2}\partial_{\mu}X_1\partial^{\mu}X_1
+\frac{1}{2}\mu_{1}^{2}X_{1}^{2} \nonumber \\
&+& \frac{1}{2}\partial_{\mu}X_2\partial^{\mu}X_2
+\frac{1}{2}\mu_{2}^{2}X_{2}^{2} \nonumber \\
&-& \lambda_{\Phi X}(\Phi^\dagger \Phi)(X_1+\alpha X_2)^{2}
-\lambda_{X_1 X_2}(X_1+\beta X_2)^{4} \nonumber \\
&+& {\cal L_{SM}} \;, 
\label{L"}
\end{eqnarray}
where the Higgs sector in the $ {\cal L_{SM}} $ is
\begin{eqnarray}
{\cal L_{SM}} &\supset& (D_{\mu}\Phi)^{\dagger}(D^{\mu}\Phi)
 +\mu^{2}(\Phi^{\dagger}\Phi)-\lambda (\Phi^{\dagger}\Phi)^{2} \;.
\end{eqnarray}
After the electroweak symmetry breaking (EWSB), the SM Higgs 
doublet field $\Phi$ is expanded around its vacuum-expectation value:
\begin{equation}
\Phi (x) = \frac{1}{\sqrt{2}} \left( \begin{array}{c}
           0 \\
           \langle \phi \rangle + \phi(x) \end{array} \right ) \;
\end{equation}
where $\langle\phi\rangle\approx 246$ GeV. Both $X_1$ and $X_2$ are also 
expanded around their vacuum-expectation values $ \langle\chi_{1/2}\rangle $:
\begin{equation}
X_{1/2}(x) = \langle\chi_{1/2}\rangle + \chi_{1/2}(x) \;
\end{equation}
Three tadpole conditions can be written down using 
$ \partial V/ \partial\phi = 0 $, 
$ \partial V/ \partial\chi_1 = 0 $, and $ \partial V/ \partial\chi_2 = 0 $,
 where $V$ is the scalar potential part of Eq.~(\ref{L"}):
\begin{equation}
\langle\phi\rangle^{2} =\frac{\mu^{2}-\lambda_{\Phi X}(\langle\chi_{1}\rangle +\alpha\langle\chi_{2}\rangle)^{2}}{\lambda} \;, 
\end{equation}
\begin{equation}
\langle\chi_1\rangle^{2} = \frac{\mu_{1}^{2}\mu_{2}^{2}-\lambda_{\Phi X}(\alpha^{2}\mu_{1}^{2}-\mu_{2}^{2})\langle\phi\rangle^{2}}{\lambda_{X_1 X_2}(\mu_{2}^{2}+\beta\mu_{1}^{2}-(\alpha -\beta)^{2}\lambda_{\Phi X}\langle\phi\rangle^{2})^{3}}\cdot (\mu_{2}^{2}-\lambda_{\Phi X}\alpha (\alpha -\beta)\langle\phi\rangle^{2})^{2} \;,
\end{equation}
\begin{equation}
\langle\chi_2\rangle^{2} = \frac{\mu_{1}^{2}\mu_{2}^{2}-\lambda_{\Phi X}(\alpha^{2}\mu_{1}^{2}-\mu_{2}^{2})\langle\phi\rangle^{2}}{\lambda_{X_1 X_2}(\mu_{2}^{2}+\beta\mu_{1}^{2}-(\alpha -\beta)^{2}\lambda_{\Phi X}\langle\phi\rangle^{2})^{3}}\cdot (\beta\mu_{1}^{2}+\lambda_{\Phi X}(\alpha -\beta)\langle\phi\rangle^{2})^{2} \;
\end{equation}
Taking the decoupling limit $ \lambda_{\Phi X}\rightarrow 0 $ from the
above equations, we recover the SM condition of $
\langle\phi\rangle^{2}=\mu^{2}/ \lambda $ as well as $
\langle\chi_1\rangle^{2}=\frac{\mu_{1}^{2}}{\lambda_{X_1 X_2}[1+\beta
    (\frac{\mu_1}{\mu_2})^{2}]^{3}} $ and $
\langle\chi_2\rangle^{2}=\frac{\beta^{2}\mu_{2}^{2}}{\lambda_{X_1
    X_2}[\beta +(\frac{\mu_2}{\mu_1})^{2}]^{3}} $.

It is easy to see that the Higgs boson $\phi$ will mix with these two
new scalar bosons $\chi_{1}$ and $\chi_{2}$ to form mass eigenstates 
denoted by $h$, $h_{D_1}$ and $h_{D_2}$,
respectively. The mass terms for the Higgs boson and these two new scalar 
bosons are
\begin{equation}
{\cal L}_m =  - \frac{1}{2} \left( \phi \; \chi _{1} \; \chi _{2} \right )\,
 \left( \begin{array}{ccc} 
        2\lambda\langle\phi\rangle^{2} & 2\lambda_{\Phi X}\langle\phi\rangle\langle\chi_{\alpha}\rangle & 2\lambda_{\Phi X}\alpha\langle\phi\rangle\langle\chi_{\alpha}\rangle \\
        2\lambda_{\Phi X}\langle\phi\rangle\langle\chi_{\alpha}\rangle & -\mu_{1}^{2} + 12\lambda_{X_1 X_2}\langle\chi_{\beta}\rangle^{2} & 12\lambda_{X_1 X_2}\beta\langle\chi_{\beta}\rangle^{2} \\
        2\lambda_{\Phi X}\alpha\langle\phi\rangle\langle\chi_{\alpha}\rangle & 12\lambda_{X_1 X_2}\beta\langle\chi_{\beta}\rangle^{2} & -\mu_{2}^{2} + 12\lambda_{X_1 X_2}\beta^{2}\langle\chi_{\beta}\rangle^{2}                                                                         \end{array} \right )\,
  \left( \begin{array}{c} 
           \phi \\
           \chi _{1} \\
           \chi _{2} \end{array} \right ) \;,
\end{equation}
where we set $ \langle\chi_{1}\rangle +\alpha\langle\chi_{2}\rangle\equiv\langle\chi_{\alpha}\rangle $ and $ \langle\chi_{1}\rangle +\beta\langle\chi_{2}\rangle\equiv\langle\chi_{\beta}\rangle $.
We can rotate $(\phi\; \chi_{1}\; \chi_{2})^T \longrightarrow (h \; h_{D_1}\; h_{D_2})^T$ through these angles $ \theta_{1} $, $ \theta_{2} $ and $ \theta_{3} $
\begin{equation}
 \left( \begin{array}{c}
                h \\ 
                h_{D_1} \\
                h_{D_2} \end{array} \right ) 
= 
 \left( \begin{array}{ccc}
           \cos\theta _{1} & \sin\theta _{1} & 0 \\
          -\sin\theta _{1} & \cos\theta _{1} & 0 \\
           0 & 0 & 1 \end{array}
 \right )\,
 \left( \begin{array}{ccc}
           \cos\theta _{2} & 0 & \sin\theta _{2} \\
           0 & 1 & 0 \\
          -\sin\theta _{2} & 0 & \cos\theta _{2} \end{array}
 \right )\,
 \left( \begin{array}{ccc}
           1 & 0 & 0 \\
           0 & \cos\theta _{3} & \sin\theta _{3} \\
           0 & -\sin\theta _{3} & \cos\theta _{3} \end{array}
 \right )\, 
 \left( \begin{array}{c}
                \phi \\ 
                \chi _{1} \\
                \chi _{2} \end{array} \right ) 
\end{equation}
\begin{equation}
 = 
 \left( \begin{array}{ccc}
           C_{\theta _{1}}C_{\theta _{2}} & (S_{\theta _{1}}C_{\theta _{3}}-C_{\theta _{1}}S_{\theta _{2}}S_{\theta _{3}}) & (S_{\theta _{1}}S_{\theta _{3}}+C_{\theta _{1}}S_{\theta _{2}}C_{\theta _{3}}) \\
           -S_{\theta _{1}}C_{\theta _{2}} & (C_{\theta _{1}}C_{\theta _{3}}+S_{\theta _{1}}S_{\theta _{2}}S_{\theta _{3}}) & (C_{\theta _{1}}S_{\theta _{3}}-S_{\theta _{1}}S_{\theta _{2}}C_{\theta _{3}}) \\
           -S_{\theta _{2}} & -C_{\theta _{2}}S_{\theta _{3}} & C_{\theta _{2}}C_{\theta _{3}} \end{array}
 \right )\,
 \left( \begin{array}{c}
                \phi \\ 
                \chi _{1} \\
                \chi _{2} \end{array} \right ) 
\end{equation}
where 
$ \theta_{1,2,3} $ is the mixing angle between $ \phi $ and $ \chi _{1} $,
between $ \phi $ and $ \chi _{2} $, and 
between $ \chi _{1} $ and $ \chi _{2} $, respectively.
$ C_{\theta_{i}} $ stands for cos $ \theta_{i} $ and 
$ S_{\theta_{i}} $ stands for sin $ \theta_{i} $.
If we assume both $ \chi_{1} $ and $ \chi_{2} $ mixings with $ \phi $ are 
very small ($ \theta_{1} $, $ \theta_{2} $ are very small), 
then it implies that $ \lambda_{\Phi X} $ is small compared to other parameters.
Thus, the masses of the Higgs boson $h$ and two scalar bosons $h_{D_1}$, 
$h_{D_2}$,
and the interaction governing $h \to h_{D_1} h_{D_1}$, $h \to h_{D_2} h_{D_2}$, 
$h \to h_{D_1} h_{D_2}$,and $h_{D_1} \to h_{D_2} h_{D_2}$
are given by, in terms of the parameters in Eq.~(\ref{L"}) in
the small $ \theta_{1} $, $ \theta_{2} $ limit,
\begin{eqnarray}
m_h^2 & \simeq & 2\lambda\langle\phi\rangle^{2}-[(\mu_{1}^{2}-12\lambda_{X_1 X_2}\langle\chi_{\beta}\rangle^{2})\sin^{2}\theta_{1}+(\mu_{2}^{2}-12\lambda_{X_1 X_2}\beta^{2}\langle\chi_{\beta}\rangle^{2})\sin^{2}\theta_{2}] \nonumber \\
&&+ 2\lambda_{\Phi X}\langle\phi\rangle\langle\chi_{\alpha}\rangle (\sin2\theta_{1}+\alpha \sin2\theta_{2}) \nonumber \\
&&= (125\;{\rm GeV} )^2 \nonumber \\
m_{h_{D_1}}^2 & \simeq & (-\mu_{1}^{2}+12\lambda_{X_1 X_2}\langle\chi_{\beta}\rangle^{2})cos^{2}\theta_{3}+(-\mu_{2}^{2}+12\lambda_{X_1 X_2}\beta^{2}\langle\chi_{\beta}\rangle^{2})\sin^{2}\theta_{3} \nonumber \\
&&+ 12\lambda_{X_1 X_2}\beta\langle\chi_{\beta}\rangle^{2}\sin2\theta_{3} \nonumber \\
&&+ 2\lambda\langle\phi\rangle^{2}\sin^{2}\theta_{1}-2\lambda_{\Phi X}\langle\phi\rangle\langle\chi_{\alpha}\rangle \sin2\theta_{1} \nonumber \\
m_{h_{D_2}}^2 & \simeq & (-\mu_{1}^{2}+12\lambda_{X_1 X_2}\langle\chi_{\beta}\rangle^{2})\sin^{2}\theta_{3}+(-\mu_{2}^{2}+12\lambda_{X_1 X_2}\beta^{2}\langle\chi_{\beta}\rangle^{2})cos^{2}\theta_{3} \nonumber \\
&&- 12\lambda_{X_1 X_2}\beta\langle\chi_{\beta}\rangle^{2}\sin2\theta_{3} \nonumber \\
&&+ 2\lambda\langle\phi\rangle^{2}\sin^{2}\theta_{2}-2\lambda_{\Phi X}\alpha\langle\phi\rangle\langle\chi_{\alpha}\rangle \sin2\theta_{2} \nonumber
\end{eqnarray}
\begin{eqnarray}
{\cal L}_{hh_{D_1}h_{D_1}} & \simeq & -\lambda_{\Phi X}\langle\phi\rangle hh_{D_1}h_{D_1} \label{Lh} \\
{\cal L}_{hh_{D_2}h_{D_2}} & \simeq & -\lambda_{\Phi X}\alpha^{2}\langle\phi\rangle hh_{D_2}h_{D_2} \label{Lk} \\
{\cal L}_{hh_{D_1}h_{D_2}} & \simeq & -2\lambda_{\Phi X}\alpha\langle\phi\rangle hh_{D_1}h_{D_2} \label{Lp} \\
{\cal L}_{h_{D_1}h_{D_2}h_{D_2}} & \simeq & -\frac{1}{2}[24\lambda_{X_1 X_2}\langle\chi_{\beta}\rangle (\beta^{2}cos^{3}\theta_{3}+\beta (\beta^{2}-2)\cos ^{2}\theta_{3}\sin\theta_{3} \nonumber \\
&&+ (1-2\beta^{2})\cos\theta_{3}\sin^{2}\theta_{3}+\beta\sin ^{3}\theta_{3})] h_{D_1}h_{D_2}h_{D_2} \nonumber \\
&& \equiv\frac{\mu_{HD} }{2} h_{D_1}h_{D_2}h_{D_2} \label{LHD} \;
\end{eqnarray}
Here we assume
$ m_{h_{D_1}} > 2m_{h_{D_2}} $ and $ h_{D_1} $ decays dominantly into 
$h_{D_2}h_{D_2}$, i.e. $ B(h_{D_1}\rightarrow h_{D_2}h_{D_2}) > 99\% $,
then we can use this property to pin down the decay width of $ h_{D_1}$ as
\begin{eqnarray}
\Gamma_{h_{D_1}} &=& \frac{1}{\tau_{h_{D_1}}}\approx 
\frac{\mu _{HD}^{2}}{32\pi m_{h_{D_1}}}\times 
\sqrt{1-4 \left(\frac{m_{h_{D_2}}}{m_{h_{D_1}}} \right)^{2}} \;.
\end{eqnarray}
The properties of the other scalar boson $ h_{D_2} $ are the same as
the scalar boson $ h_{s} $ in Higgs portal model-1. Thus, the 
partial widths for
$h_{D_2} \to \ell^+ \ell^-$ and $h_{D_2} \to \pi \pi$ are given by
\cite{Gunion:1989we}
\begin{eqnarray}
\Gamma ( h_{D_2} \to \ell^+ \ell^-) &=& \sin^2\theta_{2} \, 
  \frac{m_\ell^2 m_{h_{D_2}}}{ 8 \pi \langle \phi \rangle^2 } \left( 1 - 
  \frac{4 m_\ell^2}{m_{h_{D_2}}^2 } \right )^{3/2}\;, \\
\Gamma ( h_{D_2} \to \pi \pi) &=& \sin^2\theta_{2} \, 
  \frac{m_{h_{D_2}}^3}{ 216 \pi \langle \phi \rangle^2 } \,
 \left( 1 -   \frac{4 m_\pi^2}{m_{h_{D_2}}^2 } \right )^{1/2} \,
 \left( 1 +  \frac{11 m_\pi^2}{2 m_{h_{D_2}}^2 } \right )^2 \;, \\
\Gamma_{h_{D_2}} &=& \frac{1}{\tau_{h_{D_2}}} = \sum_{\ell = e, \mu} \Gamma(h_{D_2} \to \ell^+ \ell^- )
          +  \sum_{\pi= \pi^+,\pi^0} \Gamma(h_{D_2} \to \pi \pi )  \;, 
\end{eqnarray}
where we have also restricted $m_{h_{D_2}} \alt 1$ GeV.

\section{Constraints}
There are a number of existing constraints on these two Higgs-portal models.
All these constraints are quite generic for any light scalar boson,
which is originally a SM singlet but mixes with the Higgs boson and
thus can decay into SM fermions and the Higgs boson can decay into a
pair of such scalar bosons.

The first constraint comes from a global fit to the Higgs signal strengths
and it constrains the nonstandard decay of
the Higgs boson to be less than 0.94 MeV using the most current
data in Summer 2014 \cite{update}. The partial width for $h \to h_{s} h_{s} $
is 
\begin{equation}
\Gamma( h\to h_{s} h_{s} ) \simeq \frac{\langle \phi \rangle^2}{32\pi m_h} \,
  \left(\lambda_{\Phi X}\right)^2 \; < 0.94\;{\rm MeV}\;.  
\end{equation}
It gives a relation
\begin{equation}
\vert\lambda_{\Phi X}\vert < 0.014 \;.
\end{equation}
 
For the Higgs-portal model-2, we can use the same method to constrain 
various partial widths of $h \to h_{D_1} h_{D_1}$, 
$h \to h_{D_2} h_{D_2}$ and $h \to h_{D_1} h_{D_2}$ as follows.
\begin{eqnarray}
\Gamma( h\to h_{D_1} h_{D_1} ) \simeq \frac{\langle \phi \rangle^2}{32\pi m_h} \,
  \left( 2\lambda_{\Phi X} \right )^2 \; \\
\Gamma( h\to h_{D_2} h_{D_2} ) \simeq \frac{\langle \phi \rangle^2}{32\pi m_h} \,
  \left( 2\lambda_{\Phi X}\alpha^{2} \right )^2 \; \\
\Gamma( h\to h_{D_1} h_{D_2} ) \simeq \frac{\langle \phi \rangle^2}{16\pi m_h} \,
  \left( 2\lambda_{\Phi X}\alpha \right )^2 \;
\end{eqnarray}
\begin{eqnarray}
\Gamma(h\to h_{D_1}h_{D_1})+\Gamma(h\to h_{D_2}h_{D_2})+\Gamma(h\to h_{D_1}h_{D_2}) \nonumber \\
\simeq\frac{\langle \phi \rangle^2}{32\pi m_h} \,
  \left( 2\lambda_{\Phi X} \right )^2 (1+\alpha^{4}+2\alpha^{2}) \; < 0.94\;{\rm MeV} \nonumber \\
\Rightarrow \qquad \vert\lambda_{\Phi X}(1+\alpha^{2})\vert < 6.99\times 10^{-3} \;.
\end{eqnarray}
Another set of constraints come from the decays of $B$ mesons \cite{BK1,BK2} summarized
in Ref.~\cite{volkas}. 
\begin{itemize}
\item For $100\;{\rm MeV} < m_{h_{s}} < 210 \;{\rm MeV}$ the scalar 
boson $h_{s}$ 
can only decay into a pair of electron and positron but the decay length 
is so long that it leaves no track or energy within the detector.  The 
search for $B\to K + {\rm \it invisible}$ and fixed target experiments constrain $\sin^2\theta \lesssim 10^{-8}$.

\item For $210\;{\rm MeV} < m_{h_{s}} < 280 \;{\rm MeV}$ the scalar boson 
$h_{s}$ 
can decay into a pair of muons, fixed target experiments and the search for 
$B \to K \mu^+ \mu^-$ in LHCb and B factories constrain $\sin^2\theta \lesssim 10^{-10}$.

\item For $280\;{\rm MeV} < m_{h_{s}} < 360 \;{\rm MeV}$ 
the same experiments constrain $ \sin^{2}\theta \lesssim 10^{-10} $, except for 
a window between $ 10^{-8}\lesssim \sin^{2}\theta\lesssim 10^{-5} $.

\item For $360\;{\rm MeV} < m_{h_{s}} < 4.8  \;{\rm GeV}$ the experimental
search for $B\to K \mu^+ \mu^-$ in LHCb and B factories constrain 
$\sin^2\theta \times B(h_{s} \to \mu^+ \mu^- ) \lesssim 10^{-6}$.
\end{itemize}

For the constraint of $B\to K \mu^+\mu^-$, we follow Ref.~\cite{volkas} and use the formula
\begin{equation}
Br(B\rightarrow K h_{s})\times Br(h_{s}\rightarrow\mu^{+}\mu^{-})\times\int^{\pi}_{0}\frac{sin\theta d\theta}{2}(1-exp[\frac{-l_{xy}}{sin\theta}\frac{1}{\gamma\beta c\tau}]) \;
\end{equation}
where $ l_{xy} $ is the maximum reconstructed transverse decay distance from the beampipe, $ \gamma $, $ \beta c $ and $ \tau $ are the boost factor, speed, and lifetime of $ h_{s} $.
\footnote
{
Since B mesons are produced with a higher boost at LHCb than B factories, the integral factor $ \int^{\pi}_{0}\frac{sin\theta d\theta}{2}(1-exp[\frac{-l_{xy}}{sin\theta}\frac{1}{\gamma\beta c\tau}]) $ for the case of LHCb will be smaller than B factories as pointed out in Ref~\cite{Schmidt-Hoberg:2013hba}. Here we simply assume these two integral factors are similar, and the results are consistent with Ref~\cite{Schmidt-Hoberg:2013hba} within uncertainties. 
}
 
\begin{figure}[t!]
\centering
\includegraphics[width=4.5in]{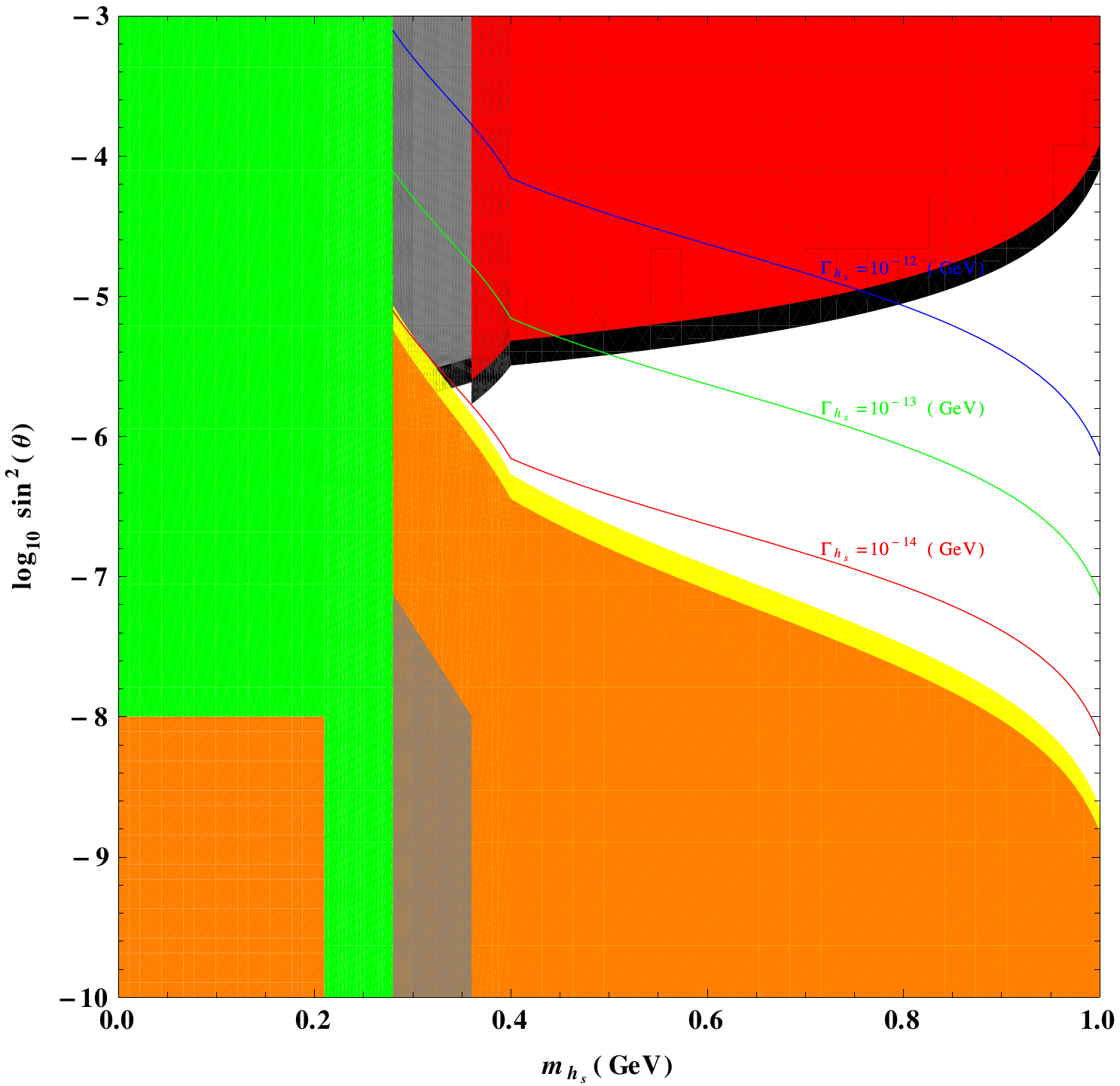}
\caption{\small \label{cons}
Existing constraints on the Higgs-portal model-1 in the plane of 
$\log_{10}\sin^2\theta$ vs $m_{h_{s}}$.
The green region is for $m_{h_s} < 280$ MeV which is ruled out by fixed target experiments, $B\to K + {\rm \it invisible}$ and $B\to K \mu^+\mu^-$. 
The gray one for $280\,{\rm MeV} < m_{h_s} < 360 \,{\rm MeV}$,
and the red one for $360\,{\rm MeV} < m_{h_s} < 1\,{\rm GeV}$, which are ruled out by fixed target experiments and $B\to K \mu^+\mu^-$ in B factories. The black region for $280\,{\rm MeV} < m_{h_s} < 1\,{\rm GeV}$ which is ruled out by $B\to K \mu^+\mu^-$ in LHCb \cite{BK1,BK2}.
The orange and yellow regions show beyond the limit of
displaced muon reconstruction for the ATLAS and CMS detector, 
respectively.
The white region then stands for the allowed parameter space
and possible muon reconstruction inside detectors.
The solid lines
are contours of various decay widths of $h_s$.
}
\end{figure}

We summarize all these constraints in Fig.~\ref{cons}.
Here we also plot beyond the limit of the displaced muon
  reconstruction for decay length $\sim 6 m$ in the ATLAS (the orange
  region) and $\sim 4 m$ in the CMS (the yellow region) for our
  analysis below.

Since the properties of the scalar boson $h_{D_2}$ in Higgs-portal 
model-2 are the same as the scalar boson $h_s$ in Higgs-portal model-1, 
we can also apply the constraints in Fig.\ref{cons} to 
$m_{h_{D_2}}$.

The third constraint, which is specific to 
the two $2\mu$-jets case,
comes from the recent search of
$h \to 2a \to  2 (\mu^+ \mu^-) +X$ by the CMS at the 8 TeV LHC \cite{cms-4u},
where $a$ is a light scalar or pseudoscalar in the mass range
of $2 m_\mu < m_a < 2 m_\tau$. The search limits at the 95\% CL is
\begin{equation}
\sigma (pp\rightarrow 2a+X)
B^{2}(a\rightarrow 2\mu)\times\epsilon_{data}\times {\cal L}
\leqslant N(m_{\mu\mu})= 3.1 + 1.2 exp(-\frac{(m_{\mu\mu}-0.32)^{2}}
{2\times 0.03^{2}}) \;,
\;
\end{equation}
where $ \epsilon_{data} $ is the experimental data efficiency, 
$ m_{\mu\mu} $ is the dimuon mass and $ {\cal L}=20.7 fb^{-1} $
\footnote
{
There are a few other similar searches \cite{cms-4u} at the LHC,
but the mass ranges are outside 1 GeV and not relevant to
 the current work.
}.
We follow closely the analysis performed in Ref.~\cite{cms-4u} at the
8 TeV run with $ \sigma (pp\rightarrow h)=19$ pb \cite{twiki}. The
branching ratio for $B(h_{s} \to \mu^+ \mu^-)$ is shown in
Table~\ref{tab:BR-hs}, and the branching ratio $B(h \to h_{s} h_{s})$
is given by $\frac{\Gamma( h \to h_{s}h_{s})}{\Gamma_h + \Gamma(h \to
  h_{s}h_{s}) }$, where $\Gamma_h \simeq 4.0 \,{\rm MeV}$. The details
of detector efficiencies will be shown in Sec. V.
However, the new light boson is restricted to decay with transverse
decay length $L_{xy} < 4.4$ cm and longitudinal decay length $L_{z} <
34.5$ cm in Ref.~\cite{cms-4u}, which are not suitable for $m_{h_{s}} \sim 0.3$ GeV in our Higgs-portal model. Finally, we found this
constraint is only applicable for $m_{h_{s}} = 0.4-0.8$
  GeV and gives the constraint $\vert\lambda_{\Phi X}\vert <
  0.007-0.026 $ in the Higgs-portal model-1. To be conservative, we
  choose $\vert\lambda_{\Phi X}\vert = 0.007$ for $m_{h_{s}} =
  0.4-1.0$ GeV in the following analysis.

Similarly, this constraint is also only applicable for
  $m_{h_{D_2}} = 0.4-0.8$ GeV and gives the constraint
  $2\vert\lambda_{\Phi X}\vert\alpha^{2} < 0.007-0.026 $ in the
  Higgs-portal model-2.


\section{Kinematical distributions for these Higgs portal models 
with benchmark points}

A lepton-jet is a very special and unique object at
colliders. In the Higgs-portal models considered in this work, 
the light scalars can decay into
leptons and pions. We focus on the 2 or 4 muons modes
in this work. 
Taking into account the constraints that
we have presented in the previous section, 
we explore the signatures for a few possible benchmark points for 
the Higgs-portal model-1 and -2, and also show the characteristics of 
$2\mu$-jets or $4\mu$-jets in the final state.

While we collect most of the kinematic distributions in appendix, here
we only illustrate the distributions which are the most relevant to the
muon-jets, namely, the angular separation among the muons within a 
muon-jet.

\subsection{Higgs-portal model-1}

\begin{figure}[t!]
\centering
\includegraphics[width=3.5in]{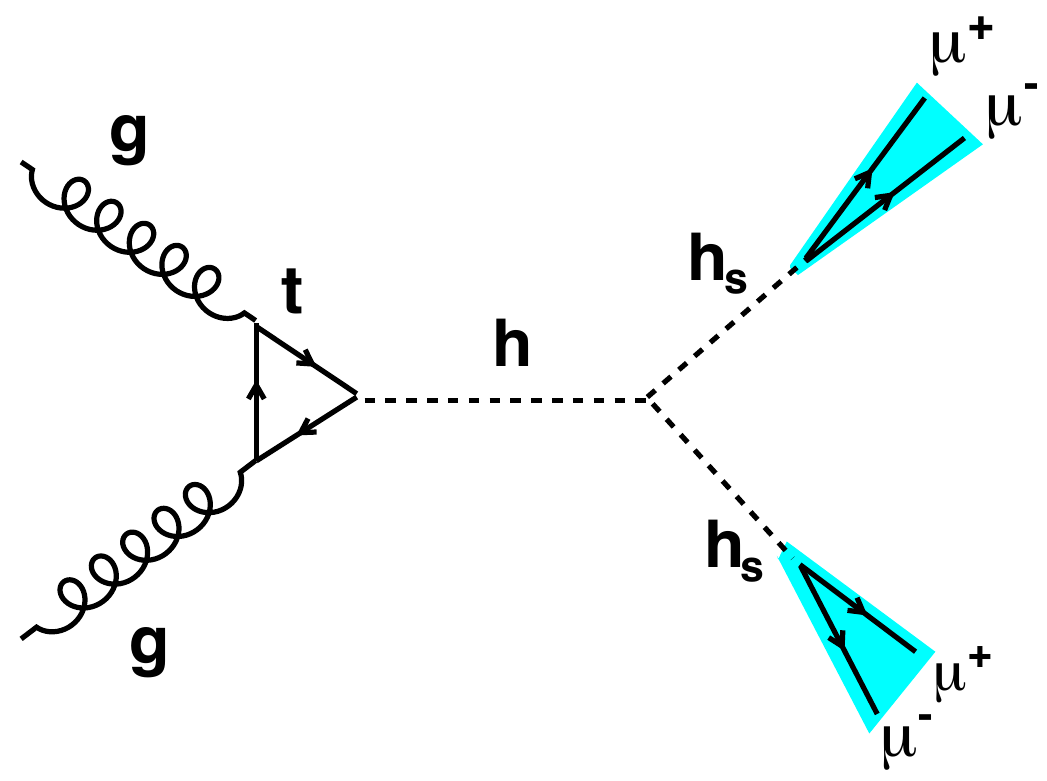}
\caption{\small \label{hs_pair}
The Feynman diagram for two $2\mu$-jets in the final state
for the Higgs-portal Model-1 (SM + one light scalar $h_s$): 
$pp \to h \to h_{s} h_{s} \to (\mu^+ \mu^- )\; (\mu^+ \mu^- )$.
}
\end{figure}

In the Higgs-portal model-1, there is only one light scalar boson in the dark
sector. 
The dominant muon-jet process comes from gluon fusion into the Higgs
boson, followed by the Higgs decay into a pair of light scalar bosons,
$h\rightarrow h_{s} h_{s}$.
Finally, each $h_{s}$ decays into a pair of opposite-sign muons. 
The Feynman diagram for this process is shown in Fig.~\ref{hs_pair}.

\begin{figure}[th!]
\centering
\includegraphics[width=3in]{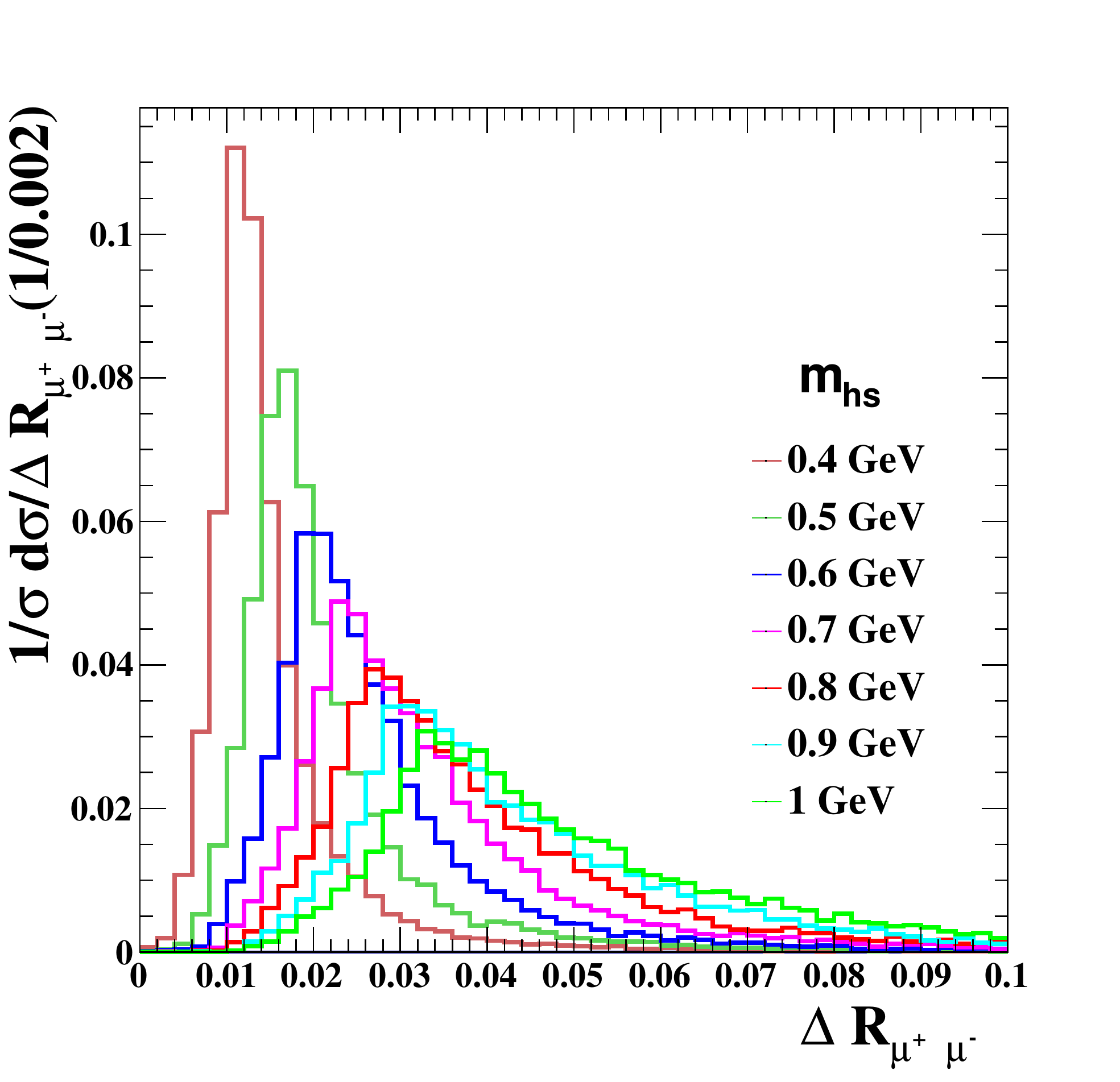}
\caption{\small \label{hs_dr} 
The opening angle $\Delta R_{\mu^+ \mu^-}$ distribution for 
a pair of oppositely-charged muons inside a muon-jet for each benchmark 
point in the Higgs-portal model-1, at LHC 14 TeV with Delphes ATLAS
simulations.
}
\end{figure}

The final state of this signal process consists of four muons, which
are organized into two dimuon pairs. Each dimuon pair consists of two
extremely collimated oppositely-charged muons.  
The angular separation
is of order $O(0.01)$.  These two dimuon pairs are back-to-back in the
transverse plane. We focus on $m_{h_{s}} =
  0.4-1.0$ GeV for varying 
$ \sin\theta $ as the benchmark points in Table~\ref{tab:BPs-hs}.
Here we also include the estimates of the lab-frame decay length
($ \gamma c\tau $) for $ h_{s} $, 
where $ \gamma c\tau\approx\frac{2}{\bigtriangleup R}\times c\tau $ 
and $ \bigtriangleup R\approx 2m_{h_{s}}/P_{T_{h_{s}}} $.
While we display most distributions in appendix, here 
we show the distribution for the opening angle $\Delta R_{\mu^+ \mu^-}$ 
for a pair oppositely-charged muons inside a muon-jet in Fig.~\ref{hs_dr}
for each benchmark point. 

\begin{figure}[h!]
\centering
\includegraphics[width=4.5in]{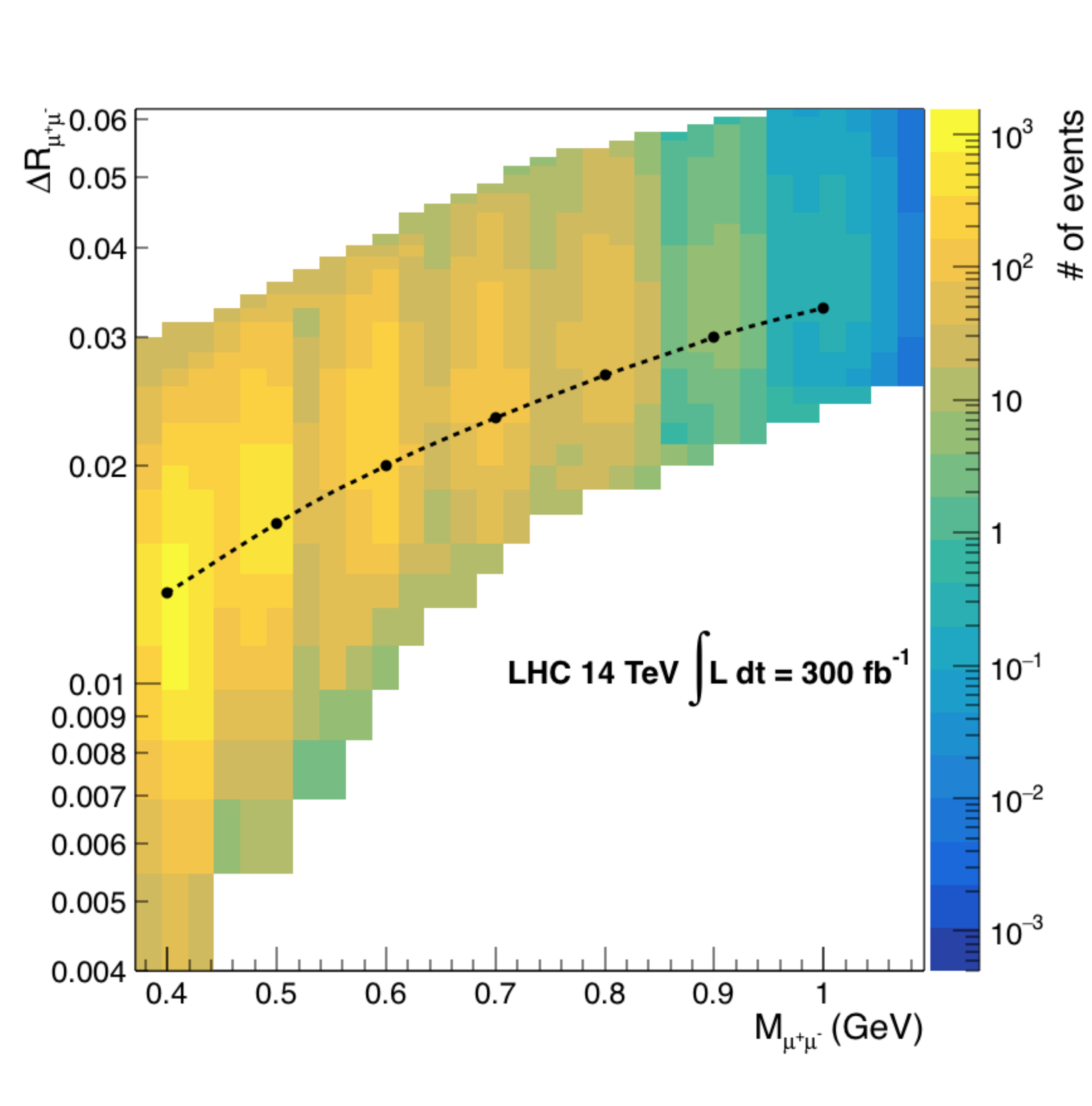}
\caption{\small \label{hs_dR_M}
The opening angle $\Delta R_{\mu^+ \mu^-}$ versus 
invariant mass $M_{\mu^+ \mu^-}$ for a pair of oppositely-charged muons
in the Higgs-portal model-1, at LHC 14 TeV, luminosity $300 fb^{-1}$ with Delphes ATLAS
simulations.
}
\end{figure}

The cross section for two $2\mu$-jets in Higgs-portal model-1 is
\begin{equation}
\sigma (pp\rightarrow h\rightarrow 2h_{s}\rightarrow 4\mu) 
 =\sigma (pp\rightarrow h)\times B(h\rightarrow h_{s}h_{s})
 \times [B(h_{s}\rightarrow \mu^{+}\mu^{-})]^{2} \;,
\end{equation}
where for 14 TeV $ \sigma (pp\rightarrow h)=49.97$ pb \cite{twiki}.
Figure~\ref{hs_dR_M} shows the distribution of $\Delta R_{\mu^+ \mu^-}$ 
versus the invariant mass of the muon pair for each benchmark point 
in Higgs-portal model-1. 
We can clearly see that the opening
angle for the dimuon pair is of order $O(0.01)$. As the mass of $h_{s}$
increases the opening angle $\Delta R_{\mu^+ \mu^-}$ 
between the two muons becomes wider, 
because the opening angle roughly scales as $m_{h_s}/ p_{T_{h_s}}$.

\begin{table}[h!]
\caption{\small  \label{tab:BPs-hs}
Signal cross sections, total decay widths, and decay lengths for 
the process $pp \to h \to 2 h_{s} \to 4\mu$
for various  benchmark points of the Higgs-portal Model-1 at LHC-14. 
We choose $m_{h_{s}} = 0.4-1.0$ GeV for various $ \sin\theta $.
Note that the innermost part of the tracker system is the pixel detector, 
which spans from 1 few cm to about 10 cm. Therefore, it can cover 
$m_{h_s} \agt 0.5$ GeV without problems. For lighter $h_s$ we can use
the outside muon spectrometer.
}
\vspace{1.0mm}
\begin{ruledtabular}
 \begin{tabular}{ l c c c c c c c }
$m_{h_{s}} $ (GeV) & 0.4 & 0.5 & 0.6 & 0.7 & 0.8 & 0.9 & 1.0 \\ \hline
$\sin\theta $ ($10^{-3}$) & 2.83 & 3.16 & 3.54 & 4.08 & 4.71 & 8.16 & 15.8 \\
$ \sigma_{14TeV}$ (fb) & 43.49 & 27.84 & 17.82 & 10.02 & 5.64 & 
0.63 & 0.04 \\
$ \Gamma_{h_{s}}$ ($10^{-13}$ GeV) & $ 1.13 $ & $ 2.69 $ & $ 5.56 $ & $ 12.2 $ & $ 25.7 $ & $ 267 $ & $ 4250 $ \\ 
$ \gamma c\tau $ (cm) & 27.3 & 9.2 & 3.7 & 1.4 & 0.6 & 0.05 & 0.003 \\ 
\hline
\end{tabular}
\end{ruledtabular}
\end{table}

\subsection{Higgs-portal model-2}

 \subsubsection{Event Topologies}

\begin{figure}[t!]
\centering
\includegraphics[width=6in]{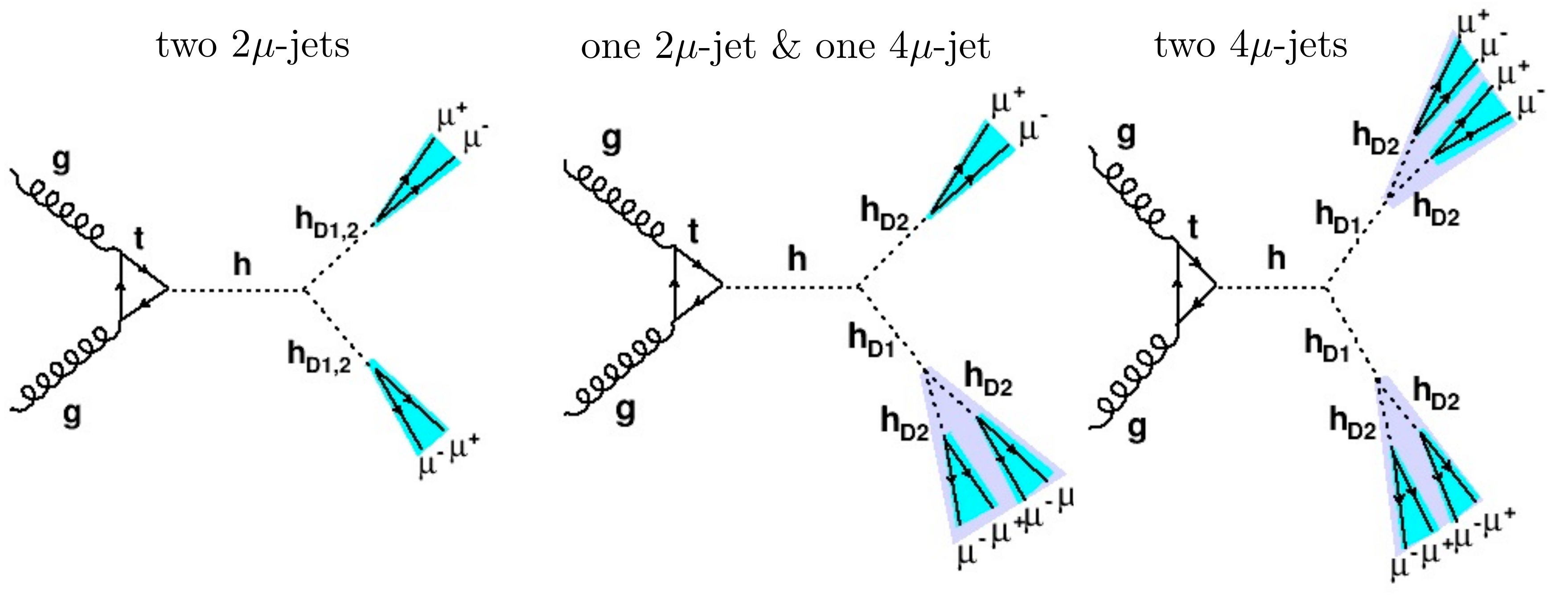} 

TP1  \hspace{1.7in}  TP2  \hspace{1.7in} TP3 
\caption{\small \label{muon-jets-234}
Feynman diagrams for muon-jet processes with the 
Higgs-portal model-2: SM + two Light Scalar $h_{D_1}$ and  $h_{D_2}$. 
Event topologies: 
(i) TP1: $ pp \to h \to h_{D1/2} h_{D1/2} \to (\mu^+ \mu^- )\;(\mu^+ \mu^- )$;
(ii) TP2: $ pp \to h \to h_{D_1} h_{D_2} \to (h_{D_2} h_{D_2}) h_{D_2} 
  \to (\mu^+ \mu^- \mu^+ \mu^- )\; (\mu^+ \mu^- )$; 
(iii) TP3: $pp \to h \to h_{D_1} h_{D_1} \to (h_{D_2} h_{D_2})\; (h_{D_2}
  h_{D_2}) \to (\mu^+ \mu^- \mu^+ \mu^- )\; (\mu^+ \mu^- \mu^+\mu^- )$.
Each pair of parentheses represents a muon-jet.
  }
\end{figure}

In the Higgs-portal model-2, there are two light scalars $h_{D_1}$ and $h_{D_2}$
in the dark sector. The dominant muon-jet processes come from gluon fusion 
into the Higgs boson, followed by the Higgs decays into two light scalars:
$h\rightarrow h_{D_1}h_{D_1}$, $h\rightarrow h_{D_1}h_{D_2}$, and
$h\rightarrow h_{D_2}h_{D_2}$. 
Here $h_{D_1}$ denotes the slightly
heavier scalar boson between the two light scalars.
The $h_{D_1}$ can decay into a pair of $h_{D_2}$, and 
we choose $h_{D_2}$ to decay into a pair of opposite-charged muons. 
Thus we can have 3 final state topologies:
\begin{enumerate}
\item  TP1: two  $2\mu$-jets, 
\item  TP2: one $2\mu$-jet and one  $4\mu$-jet, and 
\item  TP3: two  $4\mu$-jets. 
\end{enumerate}
The Feynman diagrams for these processes are shown in Fig.~\ref{muon-jets-234}.
The final states
corresponding to the event topologies TP1, TP2, and TP3
consist of 4, 6, and 8 muons, respectively, 
which are organized into two back-to-back muon-jets.
Each $2\mu$-jet is made up of a pair of oppositely-charged muons
while each $4\mu$-jet consists of two pair of oppositely-charged muons.
The angular separation between the two oppositely-charged muons 
in each $2\mu$-jet depends on the mass of the two light scalars, 
which is of order $O(0.01)$ in $\Delta R_{\mu^+\mu^-}$.
On the other hand, the angular separation between the two 
oppositely-charged muons in each $4\mu$-jet has a longer tail because
half of the times the wrong pair of muons are grouped together.

\begin{table}[th!]
\caption{\small \label{tab:BPs-case1}
Benchmark points for case 1, 2, and 3 of the Higgs-portal model-2.} 
\vspace{1.0mm}
\begin{ruledtabular}
 \begin{tabular}{lccccccccc}
$ m_{h_{D_1}} $ (GeV) & \multicolumn{6}{c}{2.5} \\
$ \sin\theta_{1} $ ($10^{-3}$) & \multicolumn{6}{c}{31.6} \\
$ \Gamma_{h_{D_1}}\, (10^{-9} $ GeV) & \multicolumn{6}{c}{4.25} \\ \hline
 & \multicolumn{2}{c}{case 1} &
                 \multicolumn{2}{c}{case 2} &
                 \multicolumn{2}{c}{case 3} \\ \hline
                 & (1) & (2) &
                   (1) & (2) &
                   (1) & (2) \\ \hline
$ m_{h_{D_2}}$ (GeV)   & 0.5 & 1.0 & 0.5 & 1.0 & 0.5 & 1.0 \\
$ \sin\theta_{2} $ ($10^{-3}$)     & 3.16 & 15.8 & 3.16 & 15.8 & 3.16 & 15.8 \\
$ \Gamma_{h_{D_2}}\, (10^{-13}$ GeV)      & 2.69 & 4250 & 2.69 & 4250 & 2.69 & 4250 \\
$ \mu_{HD}\;(10^{-3} GeV)$ & 1.08 & 1.33 & 1.08 & 1.33 & 1.08 & 1.33
\end{tabular}
\end{ruledtabular}
\end{table}

The most updated fits to the Higgs boson signal strengths
\cite{update} restrict the couplings of $ hh_{D_1}h_{D_1} $, $
hh_{D_1}h_{D_2} $ and $ hh_{D_2}h_{D_2} $ by $ \Gamma (h \rightarrow
\mbox{nonstandand} ) < 0.94$ MeV or $B(h\rightarrow \mbox{nonstandand}
) < 19\%$.
Therefore, we choose 3 different cases for different combinations of 
$ \lambda_{\Phi X} $ and $ \alpha $ as follows:
\begin{itemize}
\item case 1 : $ B(h\rightarrow h_{D_{1}}h_{D_{1}})=B(h\rightarrow h_{D_{1}}h_{D_{2}})=4\cdot B(h\rightarrow h_{D_{2}}h_{D_{2}}) $ \\
$ \lambda_{\Phi X}=4.66\times 10^{-3} $ and $ \alpha =\frac{1}{\sqrt{2}} $ ;
\item case 2 : $ B(h\rightarrow h_{D_{1}}h_{D_{1}})=10\cdot B(h\rightarrow h_{D_{1}}h_{D_{2}})=400\cdot B(h\rightarrow h_{D_{2}}h_{D_{2}}) $ \\
$ \lambda_{\Phi X}=6.65\times 10^{-3} $ and $ \alpha =\frac{1}{2\sqrt{5}} $;
\item case 3 : $ B(h\rightarrow h_{D_{1}}h_{D_{1}})=\frac{1}{10}\cdot B(h\rightarrow h_{D_{1}}h_{D_{2}})=\frac{1}{25}\cdot B(h\rightarrow h_{D_{2}}h_{D_{2}}) $ \\
$ \lambda_{\Phi X}=1.16\times 10^{-3} $ and $ \alpha =\sqrt{5} $.
\end{itemize}
We list the benchmark points for each case in
Table~\ref{tab:BPs-case1}.
We shall also display the $ p_{T} $ and $ \eta $ distributions of the 
benchmark points for case 1 with final states of 4, 6, and 8 muons in 
appendix.

\subsubsection{Simulations}

The Higgs-portal model-2 can produce 4, 6, or 8 muons in the final state 
with event
topologies TP1, TP2, and TP3.
Since the muons originate from the 125 GeV Higgs boson, 
the more the muons in the final state, the lower the transverse
momentum $p_{T_{\mu}}$ for each muon will be. Therefore, we would not get
very energetic muons in the final states with multi-muons.
The topology TP1 with two $2\mu$-jets in the final state 
suffers from the constraint of the CMS search \cite{cms-4u} just like 
the Higgs-portal model-1.
The other two topologies TP2 and TP3 containing one or more $4\mu$-jets
,
each of which is made up of four muons, and so the $p_T$ of each
muon is softer than that of each $2\mu$-jet.
At the LHC, both ATLAS and CMS experiments can
detect collimated and soft muons\cite{atlas-4u, cms-4u}.
In this work, we use the muon detection efficiency for soft muons 
(muons with $p_T <$
10 GeV) for ATLAS experiments \cite{atlas-soft-u,cms-soft-u}
in the fast detector simulation with Delphes. We use MADGRAPH v.5
\cite{madgraph} with parton showering by Pythia v.6 \cite{pythia},
detector simulations using Delphes v.3\cite{delphes3, long-live}, and the
analysis tools by MadAnalysis5 \cite{madana}.
	
The muon-jet in our Delphes simulation is defined
as~\cite{atlas-4u,Cheung:2009su}: Starting with the hardest muon we
collected all muons within $\Delta R = 0.1$ around it and added their
4-vectors to the muon-jet. This was repeated until no further muons
were found within $\Delta R = 0.1$ around the muon-jet 4-vector. This
same 4-vector was then used to define the isolation cone $0.1 < \Delta
R < 0.4$. Here we collected the muon candidates as: within a cone of
$\Delta R < 0.001$ the maximum transverse momenta sum of all charge
tracks with $P_T > 0.5\ GeV$ but the muon one is $\sum P_T <
2\ GeV$.
Then we use kinematic cuts to check if two (or four) muons will
survive the $\Delta R_{\mu^+ \mu^-} < 0.3$ (or $\Delta R_{4\mu} <
1$) cut.
\footnote{
For $4\mu-$jet reconstruction: (i) we used the angular separation of muon
pairs with $\Delta R_{\mu^+\mu^-} $ smaller than the proper cone size
($\Delta R_{\mu^+\mu^-}\sim 2 m_{hs}/P_{T_{h_s}}$ for directly decaying
$2\mu-$jet), (ii) find two oppositely charged muons within a $\Delta R $
cone with an invariant mass peaked at the lighter scalar-boson
mass to reconstruct a $2\mu-$jet, (iii) then find a pair of these $2\mu-$jets
within the $\Delta R$ cone, with an invariant mass peaked at the heavier
scalar-boson mass to reconstruct the $4\mu-$jet.  }

We are going to perform simulations for the final-state topologies of
TP1, TP2 and TP3 in case 1 of the Higgs-portal model-2.  Note that the
choice of parameters in case 1 allows all three event topologies. 
In the Higgs-portal model, the light scalars comes
from the Higgs boson decay, thus the Higgs-mass-window cut can be used to
separate the signal from backgrounds. 
We show in the appendix the invariant mass of $\mu-$jets for case 1 of 
the Higgs-portal model-2 to
illustrate the Higgs-mass window in three final-state topologies TP1,
TP2 and TP3.

\subsubsection{Angular Separation, Invariant mass and Cross Sections}

The cross sections for two $2\mu$-jets (TP1), 
one $2\mu$-jet $\&$ one $4\mu$-jet (TP2),
and two $4\mu$-jets (TP3) are given by
\begin{eqnarray}
{\rm TP1} &:& 
\sigma (pp\rightarrow h\rightarrow 2h_{D1/D2}\rightarrow 4\mu) \nonumber \\
&&  = \sigma (pp\rightarrow h)\times B(h\rightarrow h_{D1/D2}h_{D1/D2})
 \times [B(h_{D1/D2}\rightarrow \mu^{+}\mu^{-})]^{2}  \nonumber\\
{\rm TP2} &:& 
\sigma (pp\rightarrow h\rightarrow h_{D_1}h_{D_2}\rightarrow h_{D_2}h_{D_2}h_{D_2}
 \rightarrow 6\mu)   \nonumber\\
&& = \sigma (pp\rightarrow h)\times B(h\rightarrow  h_{D_1}h_{D_2})\times 
 B(h_{D_1}\rightarrow h_{D_2}h_{D_2})\times [B(h_{D_2}\rightarrow 
  \mu^{+}\mu^{-})]^{3}\nonumber\\
{\rm TP3} &:&  
\sigma (pp\rightarrow h\rightarrow h_{D_1}h_{D_1}\rightarrow 
  h_{D_2}h_{D_2}h_{D_2}h_{D_2}\rightarrow 8\mu)  \nonumber\\
&& = \sigma (pp\rightarrow h)\times B(h\rightarrow  h_{D_1}h_{D_1})\times 
  [B(h_{D_1}\rightarrow h_{D_2}h_{D_2})]^{2}\times [B(h_{D_2}\rightarrow 
  \mu^{+}\mu^{-})]^{4} \nonumber
\end{eqnarray}
The cross sections for three different event topologies for all benchmark
points are listed in Table~\ref{tab:Xsec-2}.
In all three cases of the Higgs-portal model-2, the branching ratio $B(h_{D_1}
\rightarrow h_{D_2} h_{D_2})$ is about $100\%$. 
The main difference among the three cases lies in the coupling strengths
of $h h_{D_1} h_{D_1}$, $h h_{D_1} h_{D_2}$ and $h h_{D_2} h_{D_2}$.

\begin{table}[th!]
\caption{\small \label{tab:Xsec-2}
Muon-jet cross sections at the LHC-14 for the event topologies
TP1: two $2\mu$-jets; TP2: one $2\mu$-jet \& one $4\mu$-jet; and
TP3: two $4\mu$-jets in case 1, 2, and 3.
}
\vspace{2.0mm}
\begin{ruledtabular}
 \begin{tabular}{lcccccc}
$\sigma_{14TeV}$ (fb) & \multicolumn{2}{c}{case 1} &
                 \multicolumn{2}{c}{case 2} &
                 \multicolumn{2}{c}{case 3} \\ \hline
                 & (1) & (2) &
                   (1) & (2) &
                   (1) & (2) \\ \hline
TP1              & $ 10.55 $ & $ 0.017 $ & $ 0.23 $ & $ 3.58\times 10^{-4} $ & $ 65.93 $ & $ 0.11 $ \\
TP2              & $ 4.18 $ & $ 2.67\times 10^{-4} $ & $ 0.85 $ & $ 5.45\times 10^{-5} $ & $ 2.61 $ & $ 1.67\times 10^{-4} $ \\
TP3              & $ 0.41 $ & $ 1.06\times 10^{-6} $ & $ 0.84 $ & $ 2.16\times 10^{-6} $ & $ 0.026 $ & $ 6.62\times 10^{-8} $
\end{tabular}
\end{ruledtabular}
\end{table}

\begin{figure}[th!]
\centering
\includegraphics[width=3in]{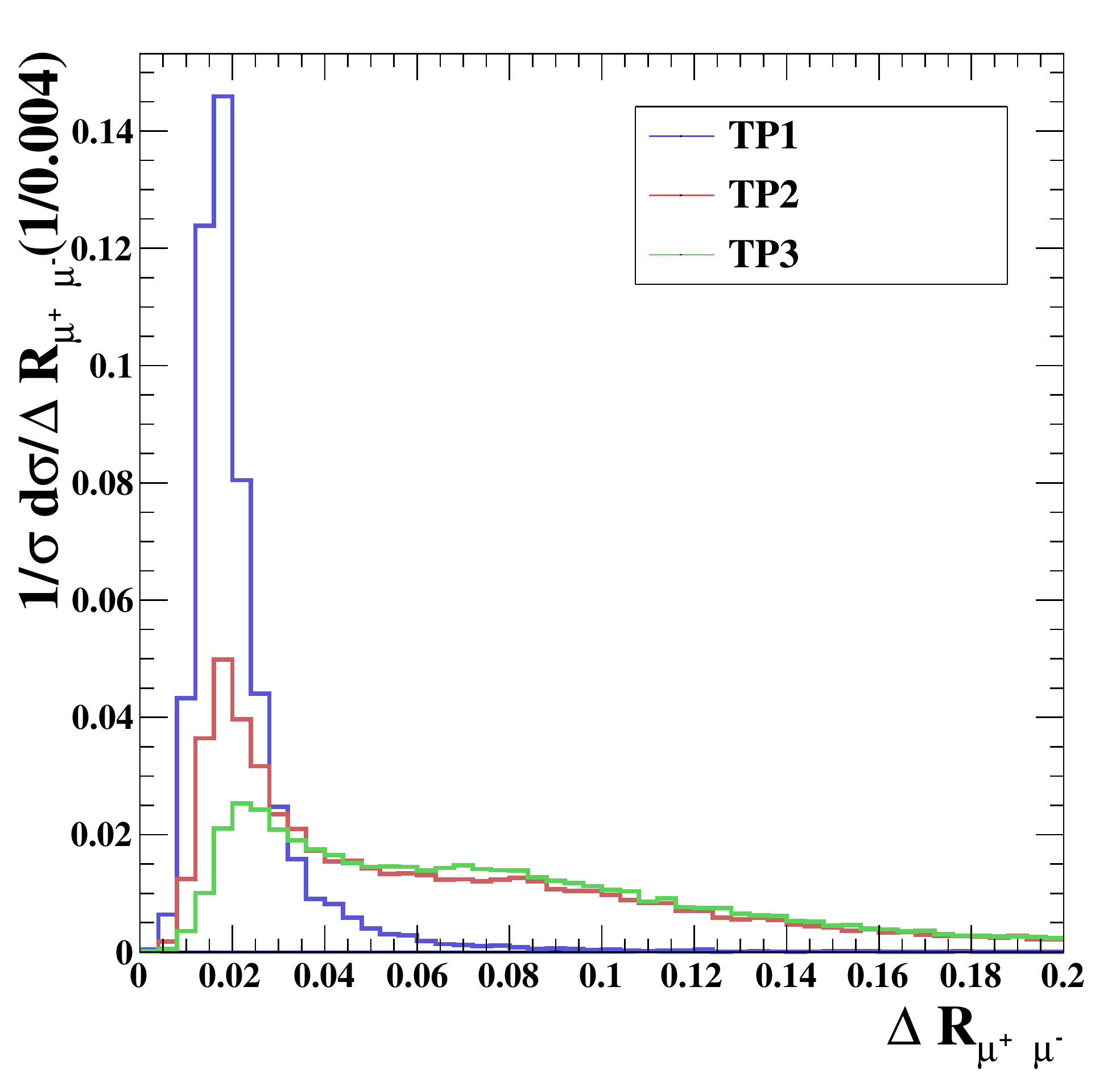}
\includegraphics[width=3in]{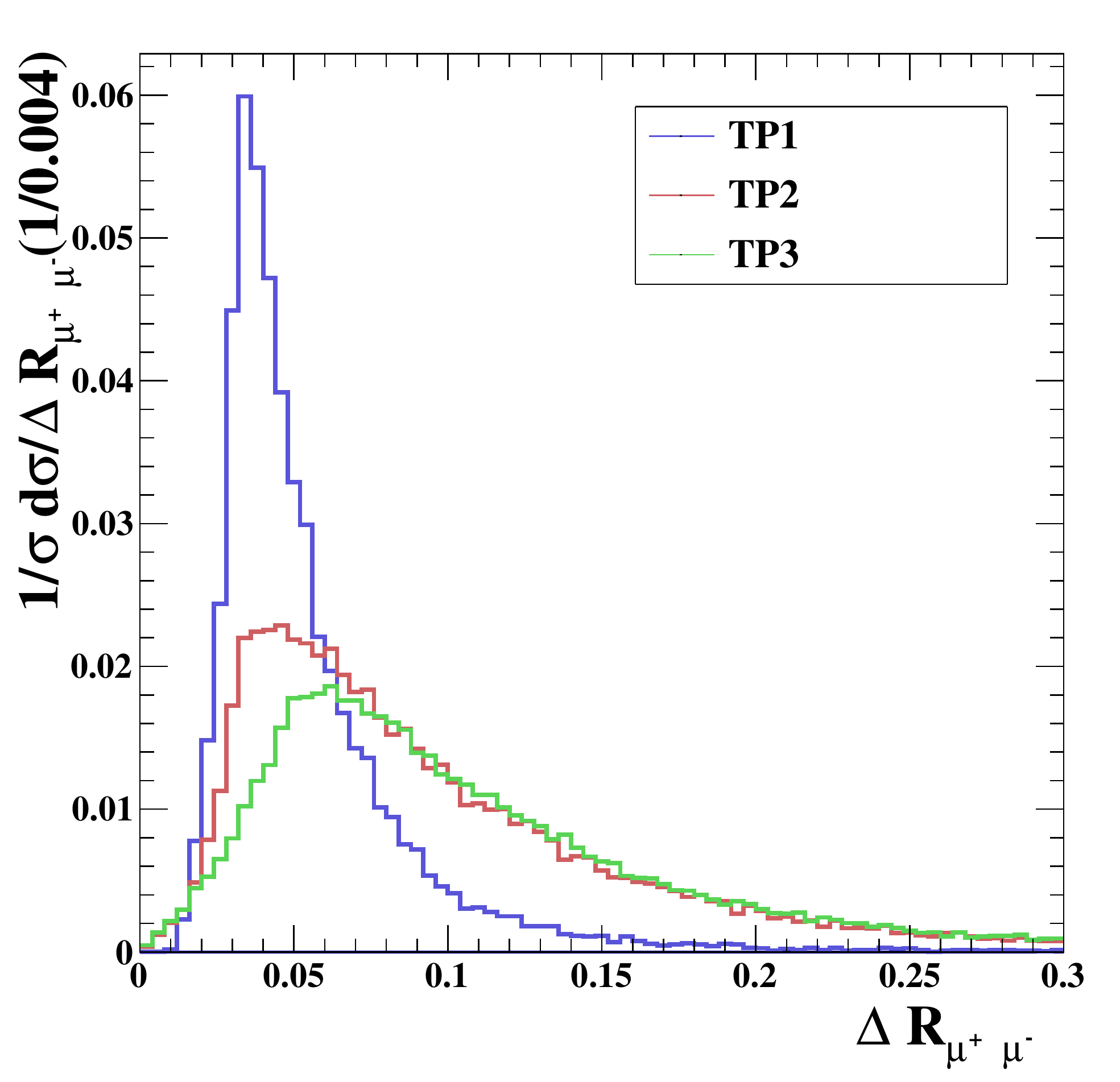}
\caption{\small \label{dR_mu}
The opening angle $\Delta R_{\mu^+ \mu^-}$ distributions for 
a pair of oppositely-charged muons inside a muon-jet. 
We show the choice of parameters for
case 1 in the Higgs-portal model-2 with $m_{h_{D_1}} = 2.5$ GeV and 
$m_{h_{D_2}} = 0.5$ (left), 1.0 GeV (right). At LHC 14 TeV with Delphes ATLAS
simulations.
}
\end{figure}

In Fig.~\ref{dR_mu}, we show the opening angle $\Delta R_{\mu^+ \mu^-}$ 
distributions for a pair of oppositely-charged muons inside a muon-jet
in different final-state event topologies TP1, TP2, and TP3.
We show the choice of parameters for 
case 1 with $m_{h_{D_1}}=2.5$ GeV and $m_{h_{D_2}}=0.5/1.0$ GeV.
We can see the cone sizes of all the TP1, TP2 and TP3 are within the order
$O(0.1)$. For each $2\mu$-jets there is only one pair of oppositely-charged 
muons in the jet cone, and so the angular separation 
$\Delta R_{\mu^+\mu^-}$ will enlarge with increases in the mass. 
For each $4\mu$-jet there are two pairs of oppositely-charged muons inside
the jet cone, and therefore the $4\mu$-jet is "fatter"' than the $2\mu$-jet.
The event topology TP2 can come from the Higgs decay into
$h_{D_1}$ and $h_{D_2}$.
We can see that the $\Delta R_{\mu^+ \mu^-}$ distribution 
has a sharp peak plus a long tail.
The sharp peak comes from the decay $h_{D_2} \rightarrow \mu^+\mu^-$,
which coincides with the first peak of TP1. On the other hand, the 
long tail comes from the decay 
$h_{D_1} \rightarrow h_{D_2}h_{D_2} \rightarrow \mu^+\mu^- \mu^+\mu^-$,
for which half of the times the wrong pair of oppositely-charged muons
are grouped together.

\section{ Sensitivity Reach at the LHC} 

The most important question is how many events for this kind of
nonstandard decays of Higgs boson that the LHC-14 with $300 fb^{-1}$
can probe via these collimated muon-jet objects. Since these
topologies in the final state  have very little background, 
we shall estimate the background event rates after applying successive
levels of cuts, and then calculate the signal event rates at 95\% CL.
For detector
efficiencies of these multimuon-jets final states, we follow
Ref. \cite{Aad:2014yea} for the non-prompt decay of light 
scalar bosons.

The major background after selection cuts dominantly comes from 
the charmonium and bottomonium production. Multiple muons can come off 
cascade semileptonic or leptonic decays, which are taken as non-prompt. 
There could easily be 4 or more muons in the final state. On the other hand, 
multiple muons which come from the low-mass Drell-Yan process 
$pp \to Z/\gamma^* \to 4 \mu$ 
and the one via Higgs boson production $pp \to h \to ZZ^* \to 4 \mu$ and 
even $ t\overline{t} $ production are taken as prompt. They are totally 
suppressed by the selection cuts.
The event rates for various backgrounds are very low. We 
shall show them momentarily.

In Ref. \cite{Aad:2014yea}, the ATLAS Collaboration 
searched for lepton-jets in the 8 TeV data with 
a luminosity of $20.3 fb^{-1}$ in two different FRVZ
models \cite{hidden}, which predict non-SM Higgs boson decays into 
lepton-jets. 
The process for the first model is
\[
  h \to f_{d_2} f_{d_2} \to (\gamma_{d}HLSP)(\gamma_{d}HLSP) \to (l^{+}l^{-})HLSP(l^{+}l^{-})HLSP \;.
\]
where $ f_{d_2} $, $ \gamma_{d} $ and $ HLSP $ are the hidden fermion, 
the dark photon and the hidden lightest stable particle in the first FRVZ model.
The final state of this model consists of two $2\mu$-jets $+$ mET.
The process for the second model is
\[
  h \to f_{d_2} f_{d_2} \to (s_{d_1}HLSP)(s_{d_1}HLSP) \to (\gamma_{d}\gamma_{d})HLSP(\gamma_{d}\gamma_{d})HLSP \;
\]
\[
\to (l^{+}l^{-})(l^{+}l^{-})HLSP(l^{+}l^{-})(l^{+}l^{-})HLSP \;.
\]
where $ s_{d_1} $ is the hidden scalar in the second FRVZ model.
The final state of this model is two $4\mu$-jets $+$ mET.

In the first model, for $ m_{\gamma_{d}}=0.4-0.9 $ GeV, the
reconstruction efficiency of muon-jets as a function of 
the transverse momentum $ p_{T} $ and the transverse 
decay distance $L_{xy}$ of
the $ \gamma_{d} $ for the $2\mu$-jet is about $ 9\%-12\% $.  In the
second model, for $ m_{s_{d_1}}=2 $GeV and $ m_{\gamma_{d}}=0.4-0.9 $
GeV, the reconstruction efficiency of muon-jets as a function of $
p_{T} $ of the $ s_{d_1} $ for the $4\mu$-jet is about $ 17\%-20\% $.
Finally, the muon trigger efficiency for $ m_{\gamma_{d}}=0.4-0.9 $
GeV as a function of $ p_{T} $ of the $ \gamma_{d} $ for $
\gamma_{d}\rightarrow\mu^{+}\mu^{-} $ is about $ 40\% $.
Note that triggering the event by seeing at least {\it one muon} is enough.

Since the final states of our Higgs-portal models are similar to
these two FRVZ models except for the mET, we will use the relevant
reconstruction efficiencies and muon trigger efficiency to simply
estimate the detector efficiencies for the non-prompt decay of 
light scalar bosons.
For the reconstruction efficiencies,  we use $ 10\% $ for the
$2\mu$-jet case and $ 20\% $ for the $4\mu$-jet case for our benchmark
points. For the muon trigger efficiency, we also use $ 40\% $ for
both  $ h_{s}\rightarrow\mu^{+}\mu^{-} $ and $
h_{D_2}\rightarrow\mu^{+}\mu^{-} $.  We summarize the detector
efficiencies for different topologies TP1, TP2 and TP3 as follows
\begin{eqnarray}
{\rm TP1} &:& 
 \epsilon\approx (10\%)^{2}\times [1-(1-(40\%))^{4}] = 8.7\times 10^{-3}  
 \nonumber\\
{\rm TP2} &:& 
 \epsilon\approx (10\%)\times (20\%)\times [1-(1-(40\%))^{6}] = 0.019 
   \nonumber\\
{\rm TP3} &:&  
 \epsilon\approx (20\%)^{2}\times [1-(1-(40\%))^{8}] = 0.039  \nonumber
\end{eqnarray}

We first look at the Higgs-portal model-1 with only one light scalar $h_s$.  
The scalar $h_s$ can decay into a pair of collimated muons.  Therefore, 
the final state consists of two $2\mu$-jets,
corresponding to the topology TP1. 
We show the observable events in Table~\ref{events-model1} for 
benchmark points in Table~\ref{tab:BPs-hs} at LHC-14 with $ 300 fb^{-1} $.
The number of events decreases gradually from 114 at $m_{h_s} = 0.4$ GeV 
down to 2 at $m_{h_s}=0.9$ GeV, which is mainly because of the decrease in
branching ratio $B(h_s \to \mu^+ \mu^-)$ (see Table~\ref{tab:BR-hs}).
Note that the decay lengths of $h_s$ for $m_{h_s} =0.4 - 0.9$ GeV are
longer than the criterion of prompt decay length (0.15 mm), and so
we use detector efficiencies of non-prompt decay for $ m_{h_s}= 0.4- 0.9$ GeV.
However, for $m_{h_s}= 1.0$ GeV the decay length is shorter than 0.15 mm and 
thus considered prompt decay, and we use the efficiencies for prompt decays.
\footnote
{ Since the efficiencies for the decay length around 0.15
    mm between non-prompt and prompt decays are a complicated continuous
    function, here we just want to simply show the major differences
    of numerical values between these two kinds of efficiencies.  }
The number of events rises to 6 for $m_{h_s}= 1.0$ GeV.
\footnote
{
We used the same selection cuts as in Table~\ref{charmon} for calculation of 
the efficiencies for prompt decays with Delphes, and got $ \epsilon = 0.460 $.  
}
\begin{table}[h!]
\caption{\small  \label{events-model1}
Number of events for the process $pp \to h \to 2 h_{s} \to 4\mu$ of the 
Higgs-portal model-1 at LHC-14 with $ 300 fb^{-1} $ for benchmark points 
in Table~\ref{tab:BPs-hs}.
}
\vspace{1.0mm}
\begin{ruledtabular}
 \begin{tabular}{ l c c c c c c c }
$ m_{h_{s}} $ (GeV) & 0.4 & 0.5 & 0.6 & 0.7 & 0.8 & 0.9 & 1.0 \\ \hline
$ \# $ of events & 114 & 73 & 47 & 26 & 15 & 2 & 6
\end{tabular}
\end{ruledtabular}
\end{table}

\begin{figure}[th!]
\centering
\includegraphics[width=3.5in]{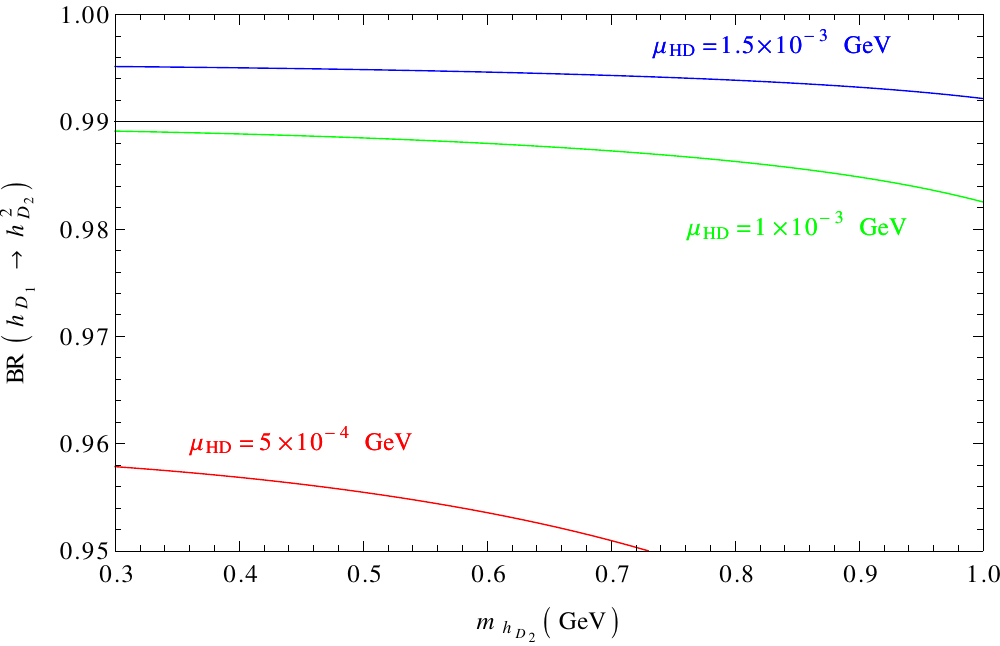}
\caption{ \small \label{m-hd2}
The branching ratio $B(h_{D_1} \to h_{D_2} h_{D_2}) $ versus $m_{h_{D_2}}$
for various $\mu_{HD}$ with fixed $m_{h_{D_{2}}}=2.5$ GeV. The purpose of 
the graph is to show how large $\mu_{HD}$ is required to give a 
branching ratio larger than $0.99$ for $m_{h_{D_{2}}}=2.5$ GeV.
}
\end{figure}

Next we consider the Higgs-portal model-2. Since the number of parameters 
involved are many, we first fix $\mu_{HD}$ which controls the branching ratio 
$B( h_{D_1}\to h_{D_2} h_{D_2} )$.  The branching ratio 
$B( h_{D_1}\to h_{D_2} h_{D_2} )$ is shown in Fig.~\ref{m-hd2} for 
fixed $m_{h_{D_{2}}}=2.5$ GeV with various values of $\mu_{\rm HD}$. For 
$\mu_{HD} = (1.0-1.5) \times 10^{-3}$ the branching ratio is almost above
$0.99$ in the mass range shown.  

The topology TP2 (one $2\mu$-jet and one $4\mu$-jet) and topology TP3
(two $4\mu$-jets) can come from case 1, 2 and 3 of the Higgs-portal
model-2. We show the observable events in
Table~\ref{events-model2} for benchmark points in
Table~\ref{tab:BPs-case1} at LHC-14 with $ 300 fb^{-1} $. Here we only
show number of events larger than 1.
\footnote
{ Similarly, We also used the same selection cuts as in
  Table~\ref{charmon} for calculation of the efficiencies for prompt
  decays with Delphes to TP1, TP2, and TP3, and got $ \epsilon = 0.46
  $ for TP1, $ \epsilon = 0.194 $ for TP2, and $ \epsilon = 0.145 $
  for TP3.  }

\begin{table}[th!]
\caption{\small \label{events-model2}
Number of events for TP1, TP2 and TP3 of Higgs portal model-2 at LHC-14 with $ 300 fb^{-1} $ for benchmark points in Table~\ref{tab:BPs-case1}.
}
\vspace{2.0mm}
\begin{ruledtabular}
 \begin{tabular}{lcccccc}
$ \# $ of events & \multicolumn{2}{c}{case 1} &
                 \multicolumn{2}{c}{case 2} &
                 \multicolumn{2}{c}{case 3} \\ \hline
                 & (1) & (2) &
                   (1) & (2) &
                   (1) & (2) \\ \hline
TP1              & 28 & 2 & 1 & -- & 172 & 15 \\
TP2              & 24 & -- & 5 & -- & 15 & -- \\
TP3              & 5 & -- & 10 & -- & -- & --
\end{tabular}
\end{ruledtabular}
\end{table}

\begin{table}[th!]
\caption{\small \label{charmon}
Number of background events for TP1, TP2 and TP3 at LHC-14 with $ 300 fb^{-1} $.}
\vspace{2.0mm}
\begin{ruledtabular}
 \begin{tabular}{lcccc}
Cuts/$ \# $ of BG events & TP1 & TP2&TP3\\
$N(\mu)=4, (6,8)$& 485452&236522&82104\\
$p_{T_\mu}>5 GeV$&50667&34724&15138\\
$|\eta_\mu|<2.4$&50667&34724&15138\\
$p_{T_{\mu_1}}>20 GeV$&23873	&	17441	&	7936\\
$115GeV<|M_{\sum\mu_i}|<135 GeV$& 28&	59&	28\\
$M_{2\mu}<3 GeV$, $\Delta R_{2\mu}<0.3$&8.49&	21.22&	\\
$M_{4\mu}<3 GeV$, $\Delta R_{4\mu}<1$&&&27.59\\
\end{tabular}
\end{ruledtabular}
\end{table}

We perform background calculations for $4,6,8$ muons to form 
muon-jets under successive cuts. The charmonium and bottomonium are 
the dominant backgrounds. We start with $3.18\times 10^6$ events 
(corresponding to the background cross section with 300 fb$^{-1}$), 
and show the 
subsequent numbers after each level of cuts in Table~\ref{charmon}.
At the end of the cut flow, the number of background events remaining
are $8.49, 21.22$ and $27.59$ for TP1, TP2, and TP3, respectively.
Thus, the 95\%CL upper limits (roughly $Z=2$)
\footnote
{
The signal significance $Z$ defined as
\begin{equation}
Z= \sqrt{2\cdot ((s+b)\cdot ln(1+s/b)-s)} \;,
\end{equation}
where $s$ and $b$ are the expected number of signal and background events, 
respectively.
}
for signal event numbers are
$6.46, 9.86,11.15$, respectively.  We then use these signal event rates
to show the sensitivity reach in the parameter space.

We can now compare the sensitivity reach by the topologies TP1, TP2, and TP3.
The more muons to be seen,
the higher the price has to be paid for detection efficiency. Nevertheless,
the signature of one $2\mu$-jet and one $4\mu$-jet in the final state
is one of the most striking decays of the Higgs boson that we can imagine. 
It implies the existence very light particles involved in the decay chain.
Similarly, two $4\mu$-jets in the final state also signal
multiple light scalar bosons in the dark sector.  

\begin{figure}[t!]
\centering
\includegraphics[width=3in]{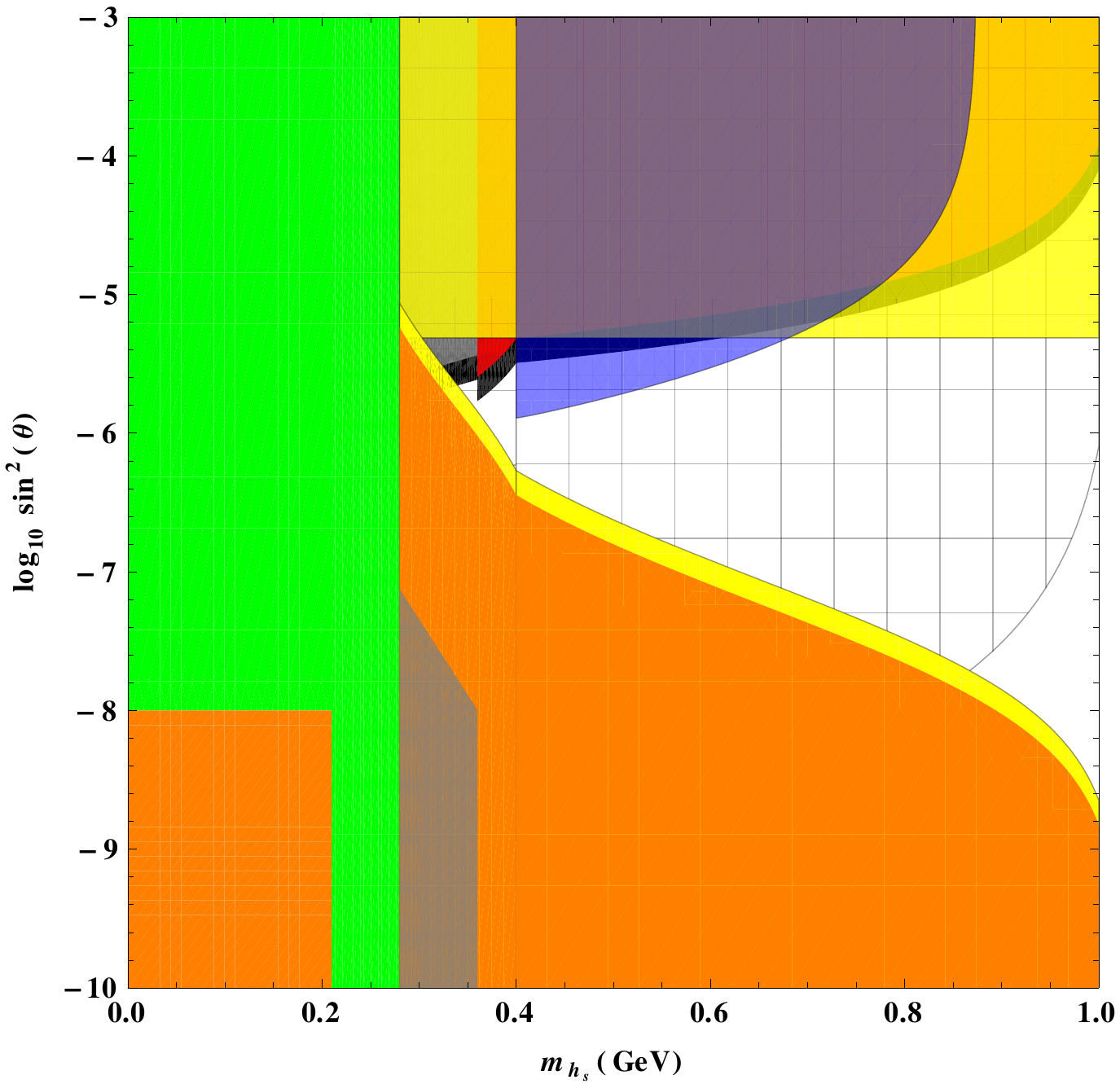}\includegraphics[width=3in]{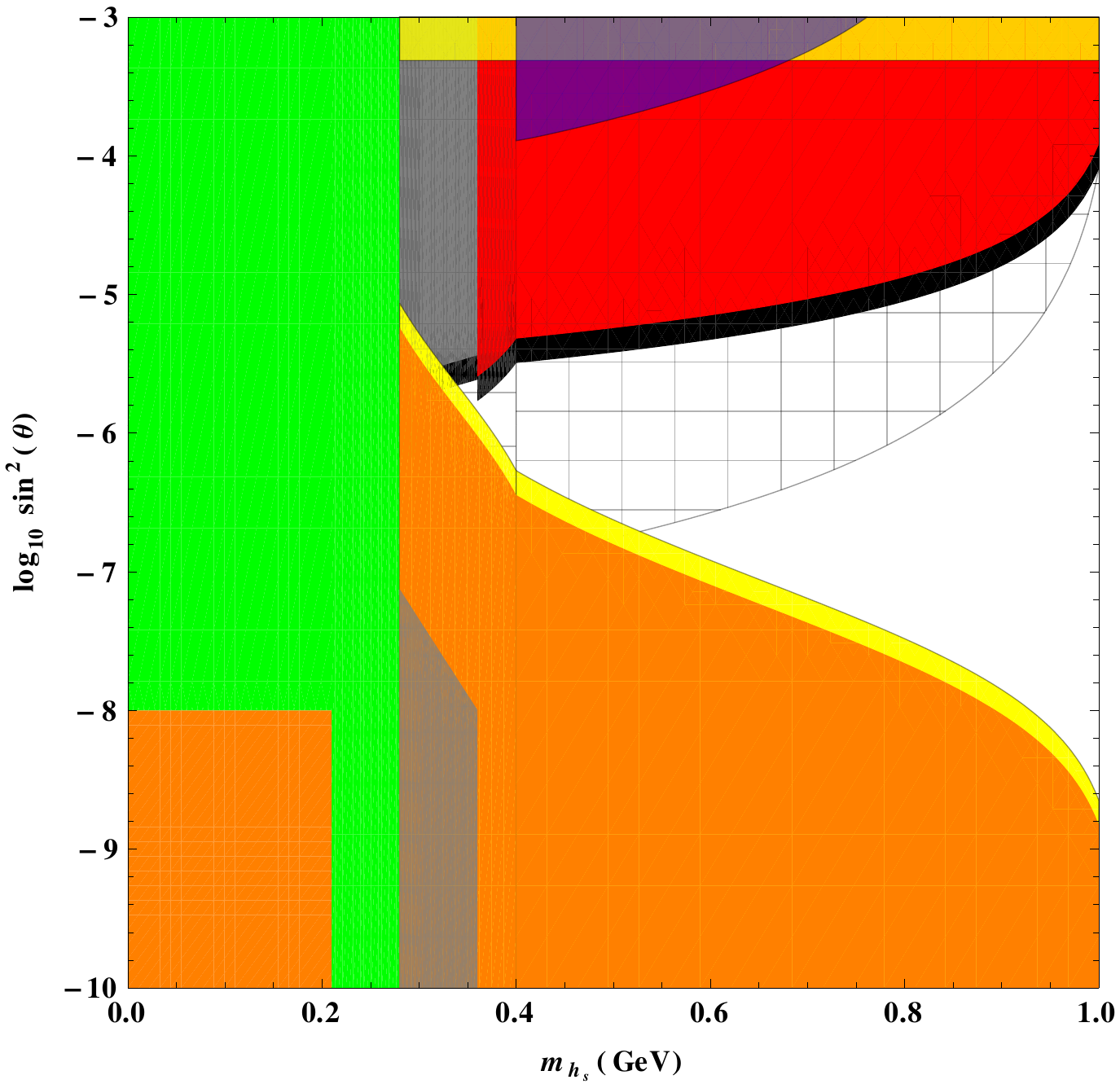}
\caption{\small \label{<X>}
Similar to Fig.~\ref{cons}, but adding the first constraint in the 
upper yellow region and third constraint in the upper blue region. 
We also use the hatched region to display the 95\%CL upper 
limits (roughly $Z=2$) in our analysis of LHC-14 with 300 fb$^{-1}$ 
to show the sensitivity reach for $ \langle\chi\rangle = 10$ GeV (left panel) 
and $ \langle\chi\rangle = 100$ GeV (right panel), respectively.
}
\end{figure}

 First, we start from the Higgs-portal model-1. 
After adding all the constraints described in Sec. III shown in 
Fig.~\ref{cons}, we can 
further use the 95\%CL upper limits (roughly $Z=2$)
in our analysis of LHC-14 with 300 fb$^{-1}$ to show the sensitivity 
reach for $ \langle\chi\rangle = 10$ GeV  and for $ \langle\chi\rangle = 100$ 
GeV in Fig.~\ref{<X>}. 
Note that in Fig.~\ref{<X>} the orange and yellow regions show
beyond the limit of the displaced muon reconstruction for decay length 
for ATLAS and CMS, respectively. The gray hatched region is where
our analysis can cover. We can see from these figures, LHC-14 with 
300 fb$^{-1}$ in our analysis could cover all the parameter space of 
$ m_{h_s} < 0.5$ GeV within possible muon reconstruction inside the detectors.

\begin{figure}[t!]
\centering
\includegraphics[width=3in]{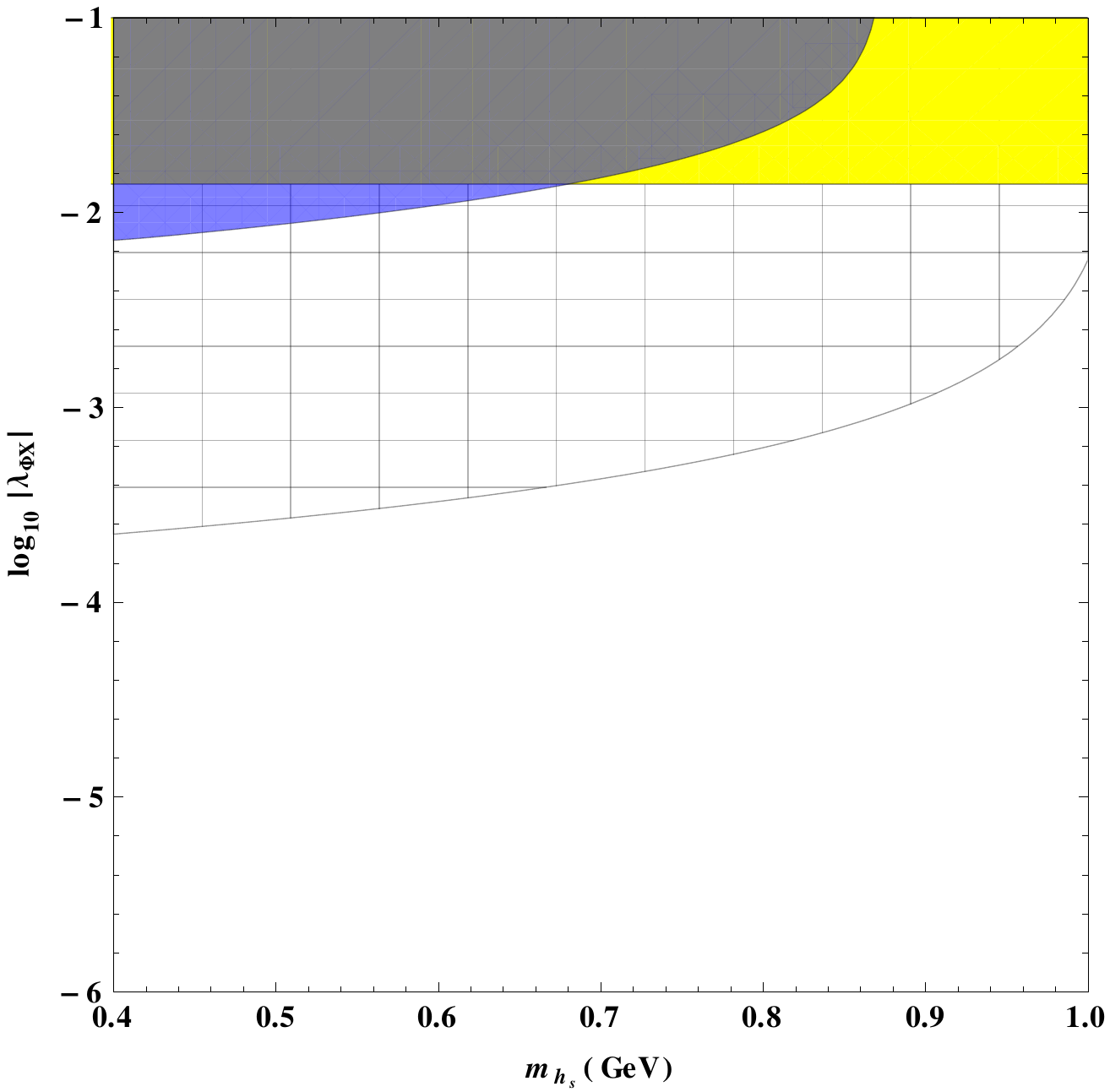}
\caption{\small \label{SP-model1}
Existing constraints and the sensitivity reach of LHC-14 with 
300 fb$^{-1}$ for the Higgs-portal model-1 in the plane of 
$\log_{10}\vert\lambda_{\Phi X}\vert$ vs $m_{h_{s}}$.
The yellow region is from the first constraint,
the blue region is from the third constraint,
and the hatched shading lines is to display the 95\%CL upper 
limits (roughly $Z=2$)
in our analysis of LHC-14 with 300 fb$^{-1}$.
}
\end{figure}

While the parameter space in the plane of $\log_{10}\sin^2\theta$ vs
$m_{h_{s}}$ for the Higgs-portal model-1 depends on the choice of $
\langle\chi\rangle $, we can also show the parameter space
in the plane of $\log_{10}\vert\lambda_{\Phi X}\vert$ vs $m_{h_{s}}$
in Fig.~\ref{SP-model1},
which is independent of the choice of $ \langle\chi\rangle $.
\footnote
{
The fundamental parameters in the Higgs-portal model-1 are $ \lambda $, $ \lambda_{X} $, $ \lambda_{\Phi X} $, $ \langle\phi\rangle $, and $
\langle\chi\rangle $. $ \theta $ can be derived from these fundamental parameters.  
}
This plot can allow us to have more direct comparison with the plots
for Higgs-portal model-2.

\begin{figure}[th!]
\centering
\includegraphics[width=3in]{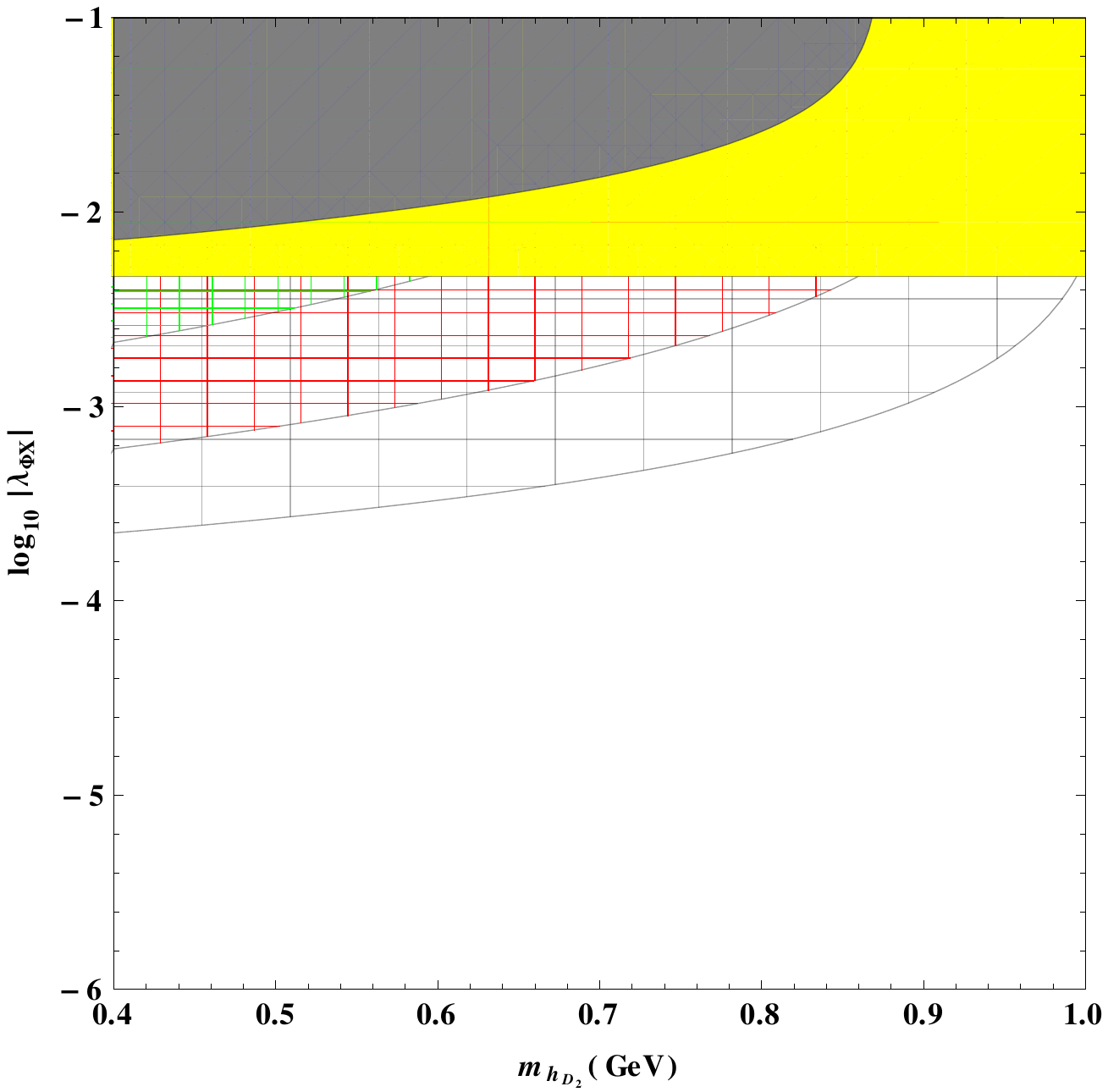}
\includegraphics[width=3in]{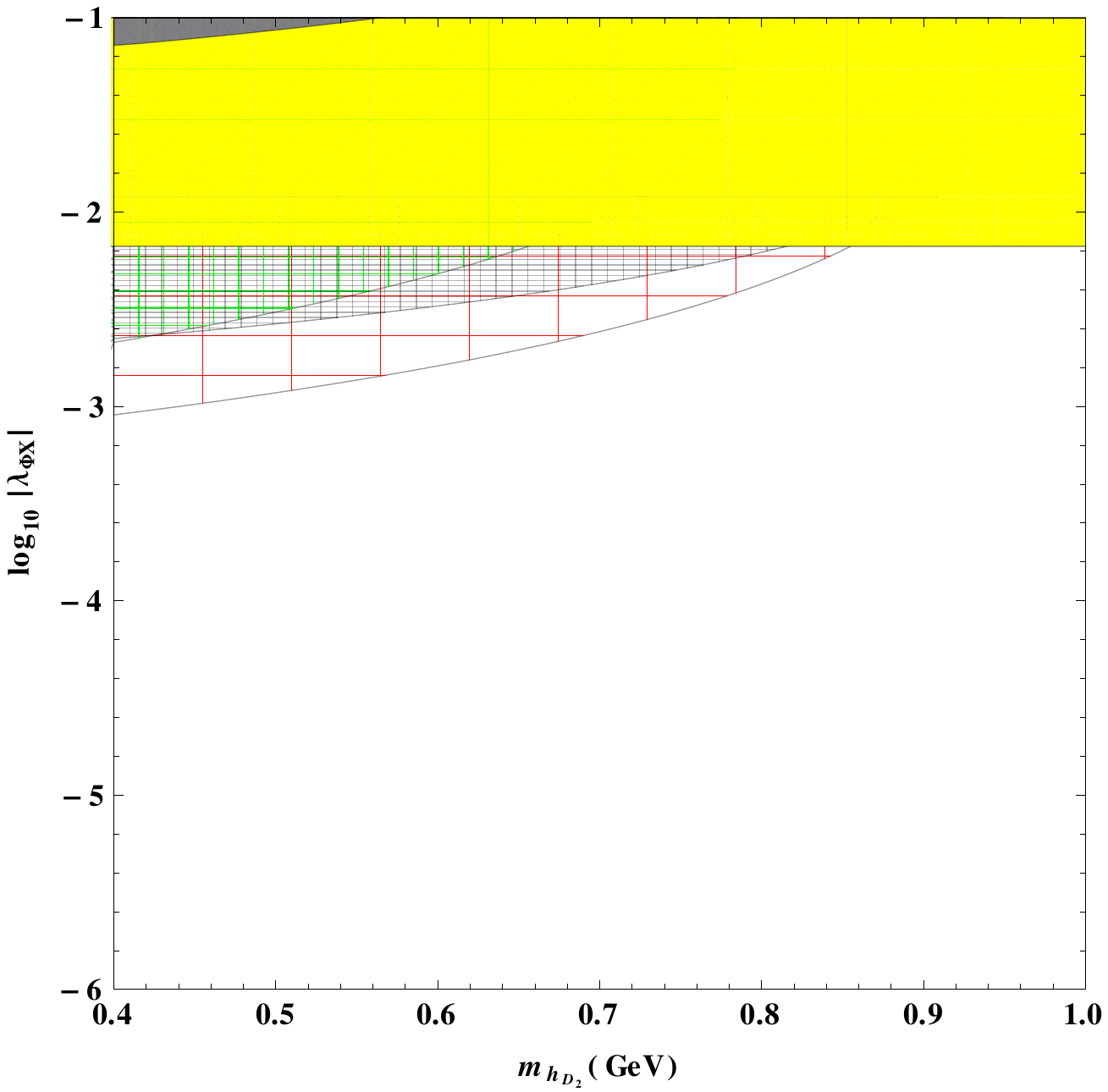}
\includegraphics[width=3in]{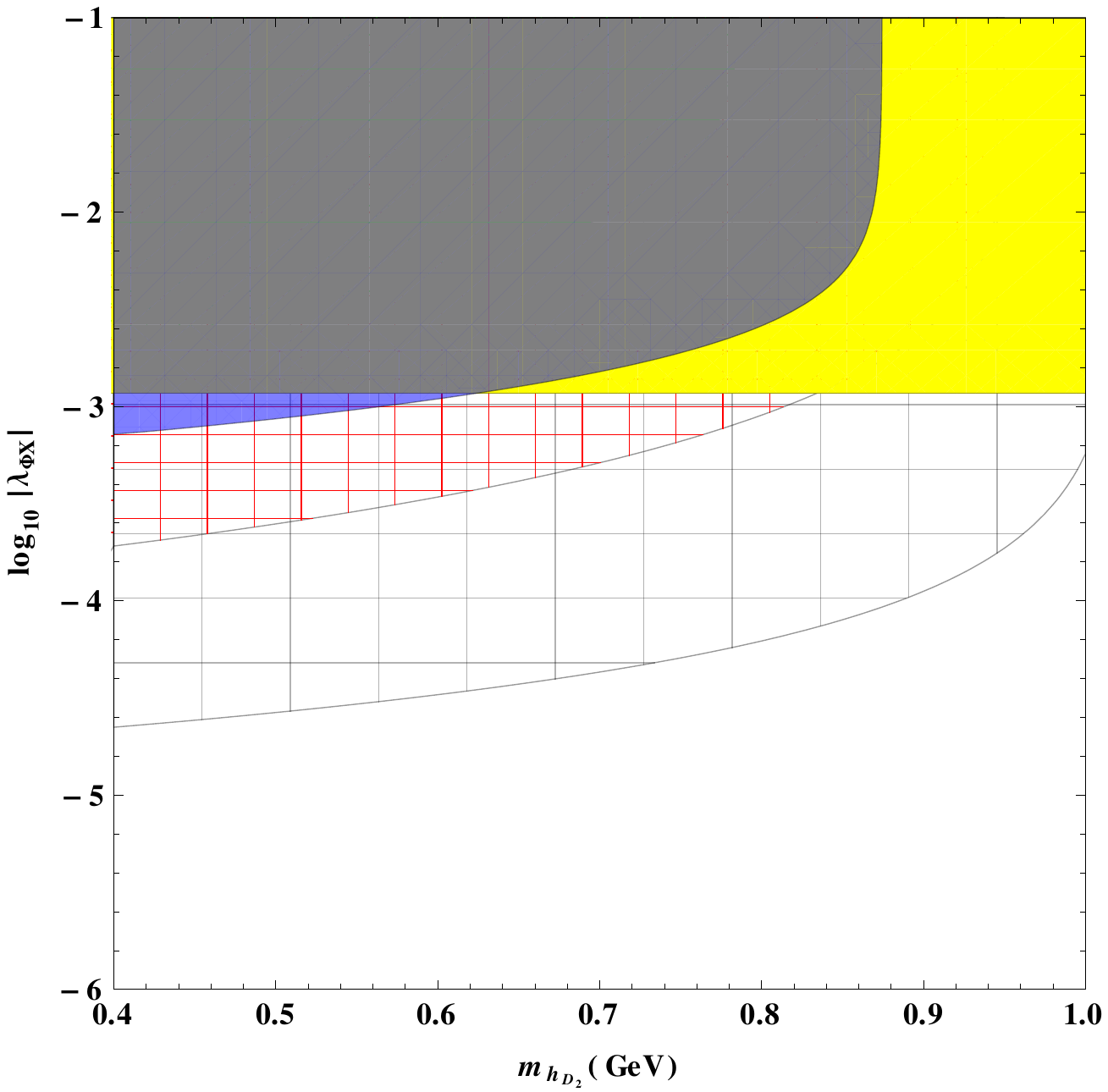}
\caption{\small \label{SP-model2}
Existing constraints and the sensitivity reach of LHC-14 with 300
fb$^{-1}$ for the Higgs-portal model-2 in the plane of
$\log_{10}\vert\lambda_{\Phi X}\vert$ vs $m_{h_{D_2}}$ for case 1
(upper left panel), case 2 (upper right panel), and case 3 (lower
panel).  The yellow region is from the first constraint, the blue
region is from the third constraint, and the hatched shading lines are
to display the 95\%CL upper limits (roughly $Z=2$) in our analysis of
LHC-14 with 300 fb$^{-1}$ for TP1 (Blue), TP2 (Red), and TP3 (Green).
}
\end{figure}

\begin{figure}[th!]
\centering
\includegraphics[width=3in]{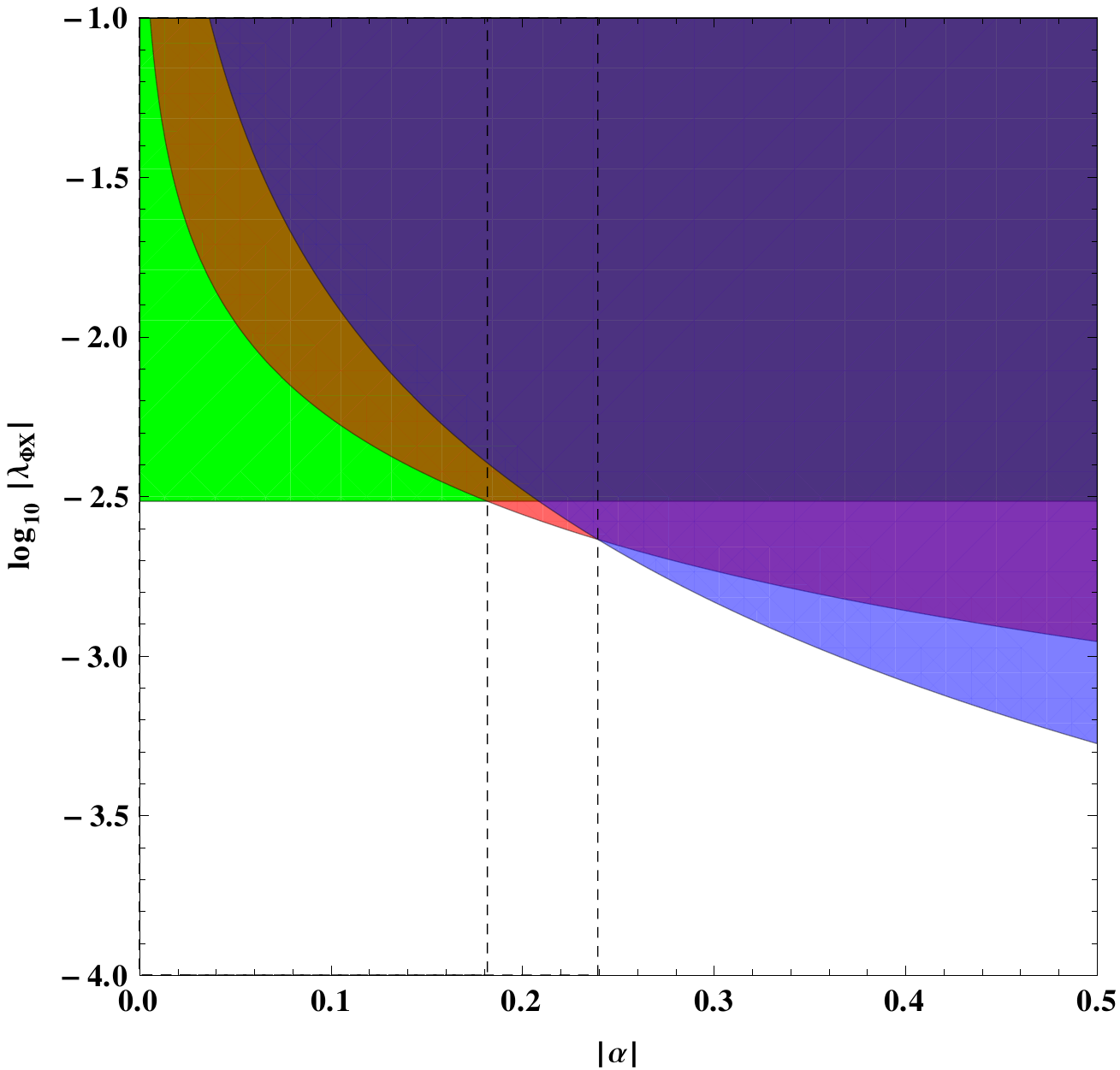}
\caption{\small \label{alpha}
The sensitivity reach of LHC-14 with 300 fb$^{-1}$ for the Higgs-portal 
model-2 in the plane of 
$\log_{10}\vert\lambda_{\Phi X}\vert$ vs $ \vert\alpha\vert $ 
with $m_{h_{D_2}}=0.5$GeV.
The color regions are to display the 95\%CL upper limits (roughly $Z=2$)
in our analysis of LHC-14 with 300 fb$^{-1}$ for TP1 (Blue), 
TP2 (Red), and TP3 (Green). 
}
\end{figure}

Next we can use the similar approach to show the parameter space in the plane
of $\log_{10}\vert\lambda_{\Phi X}\vert$ vs $m_{h_{D_2}}$ for case 1,
2 and 3 in the Higgs-portal model-2 in Fig.~\ref{SP-model2}. An
interesting observation is that there are some crossovers among different
hatched regions in the figure of case 2. To further explore this
property, we fix $m_{h_{D_2}}=0.5$GeV and vary different values of $
\vert \alpha\vert $ in the plane of $\log_{10}\vert\lambda_{\Phi
  X}\vert$ vs $ \vert \alpha\vert $ in Fig.~\ref{alpha}. We can see
when $ \vert \alpha\vert\lesssim 0.18 $ the best sensitivity reach of 
LHC-14 with 300 fb$^{-1}$ is using the TP3 topology, then in the
range of $ 0.18\lesssim\vert \alpha\vert\lesssim 0.24 $ turns out to be
TP2, finally after $ \vert \alpha\vert\gtrsim 0.24 $ the best reach is
given by TP1. 
Such a feature can also be observed for other
values of $m_{h_{D_2}}$.  
Another observation is that when
$\vert\alpha\vert$  becomes small, the constraint on
$\vert\lambda_{\Phi X}\vert$ will also be less stringent.

\section{Conclusions}
Muon-jets are interesting and clean signatures at 
colliders, provided the angular resolution of muons are fine enough
to differentiate them.
The current designs of the ATLAS and CMS have such capabilities of
probing angular separation as small as $10^{-3}$.
In general, muon-jets arise from the decay of fast-moving light particles.
In this work, we have demonstrated a couple of dark-sector models, in
which there are a number of very light scalar bosons,
which can be accessed via the Higgs boson decays.
We have investigated the signatures of $2\mu$-jets and $4\mu$-jets,
which consist of, respectively,  one and two pairs of 
oppositely-charged muons in a very narrow cone defined by 
$\Delta R \alt 0.01$.

In the Higgs-portal model-1 that we considered, the final state consists of 
two $2\mu$-jets. The current experimental search for such a final state
has put on it a tight constraint, such that the allowable cross section
becomes very small. On the other hand, in the Higgs-portal model-2 that we 
considered the final-state event topologies can have 
(i) two $2\mu$-jets (TP1), (ii) one $2\mu$-jet and one $4\mu$-jet (TP2),
or (iii) two $4\mu$-jets (TP3). Even though the topologies TP2 and TP3 
are still not yet fully constrained 
from either ATLAS or CMS, their 
allowable cross sections are yet small. We have also looked at the 
invariant mass and the angular separation of the oppositely-charged muon pair,
which show interesting features that can help distinguishing various 
topologies.  Thus, it helps to pin down the masses of the underlying 
light scalar bosons.

Before we close we offer the following comments.
\begin{enumerate}
\item 
Since the topologies TP2 and TP3 still allow sizeable cross sections
and almost background free, we encourage our experimental groups to
focus on these kind of final states and these results will let us know
more about the structure of more general dark sector.

\item
For the Higgs-portal model-2, if we use much heavier $h_{D_{1}}$, say
$m_{h_{D_{1}}}\gtrsim 10$ GeV, then we will just see some very
collimated muon pairs instead of a "fat" muon-jet without substructure
inside it. So it is more interesting to analyze both $h_{D_{1}}$ and
$h_{D_{2}}$ are of mass about $\mathcal{O}$(1 GeV).

\item
For the Higgs-portal model-2, if $m_{h_{D_{2}}} < m_{h_{D_{1}}} <
2m_{h_{D_{2}}}$, then $h_{D_{1}}$ can decay into $4\mu$-jet by one
on-shell and one off-shell $h_{D_{2}}$ which will have different
substructure inside $4\mu$-jet from the case of $ m_{h_{D_{1}}} >
2m_{h_{D_{2}}} $, but its cross section is also suppressed.

\item
In this work we just investigated the signatures of $2\mu$-jets and
$4\mu$-jets for three different final-state event topologies. However, if
we take into account the three-body decay of $ h\rightarrow
h_{D_{1/2}}h_{D_{1/2}}h_{D_{1/2}} $ and $ h_{D_{1}}\rightarrow
h_{D_{2}}h_{D_{2}}h_{D_{2}} $, then we will have more different final-state 
event topologies, including $6\mu$-jets, which can enrich the analysis
but are seriously suppressed by the phase space.

\item
Our simple models are quite generic for any more complicated models,
which include either one or more very light scalar bosons mixing with
the SM Higgs boson. There are at least one long-lived neutral particle(s)
in this kind of models, which are still testable below 1 GeV for both the
ATLAS and CMS as shown in Fig.~\ref{cons}. Therefore, we encourage our
experimental groups to perform the analysis of real detector effects
of the displaced muon reconstruction efficiency of this kind of
scenario to further confirm this possibility.

\end{enumerate}

We have demonstrated that the existence of muon-jets such as 
$2\mu$-jets or $4\mu$-jets would signal the presence of very light
scalar bosons, perhaps coming from dark sectors.
We therefore suggest our experimental colleagues to
look into the $n\mu$-jets with $n>2$. The findings of such objects
are definitely signals of new physics and help us to understand the dark 
sector connecting to the Higgs sector.

\appendix

\section{Kinematical Distributions}
Here we collect all the kinematical distributions for model-1 and model-2

\subsection{Higgs-portal model-1}

\begin{figure}[th!]
\centering
\includegraphics[width=3in]{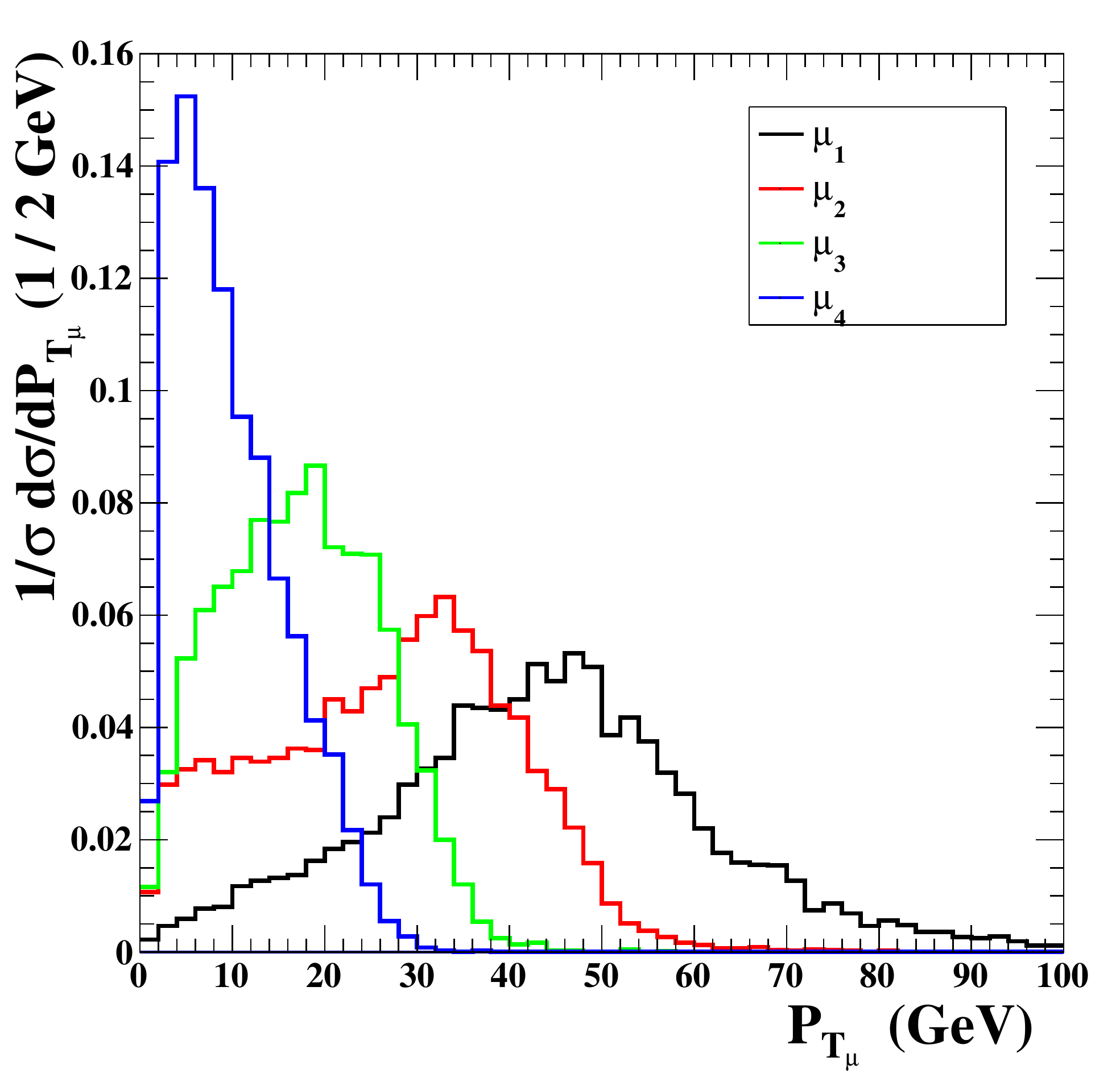}\includegraphics[width=3in]{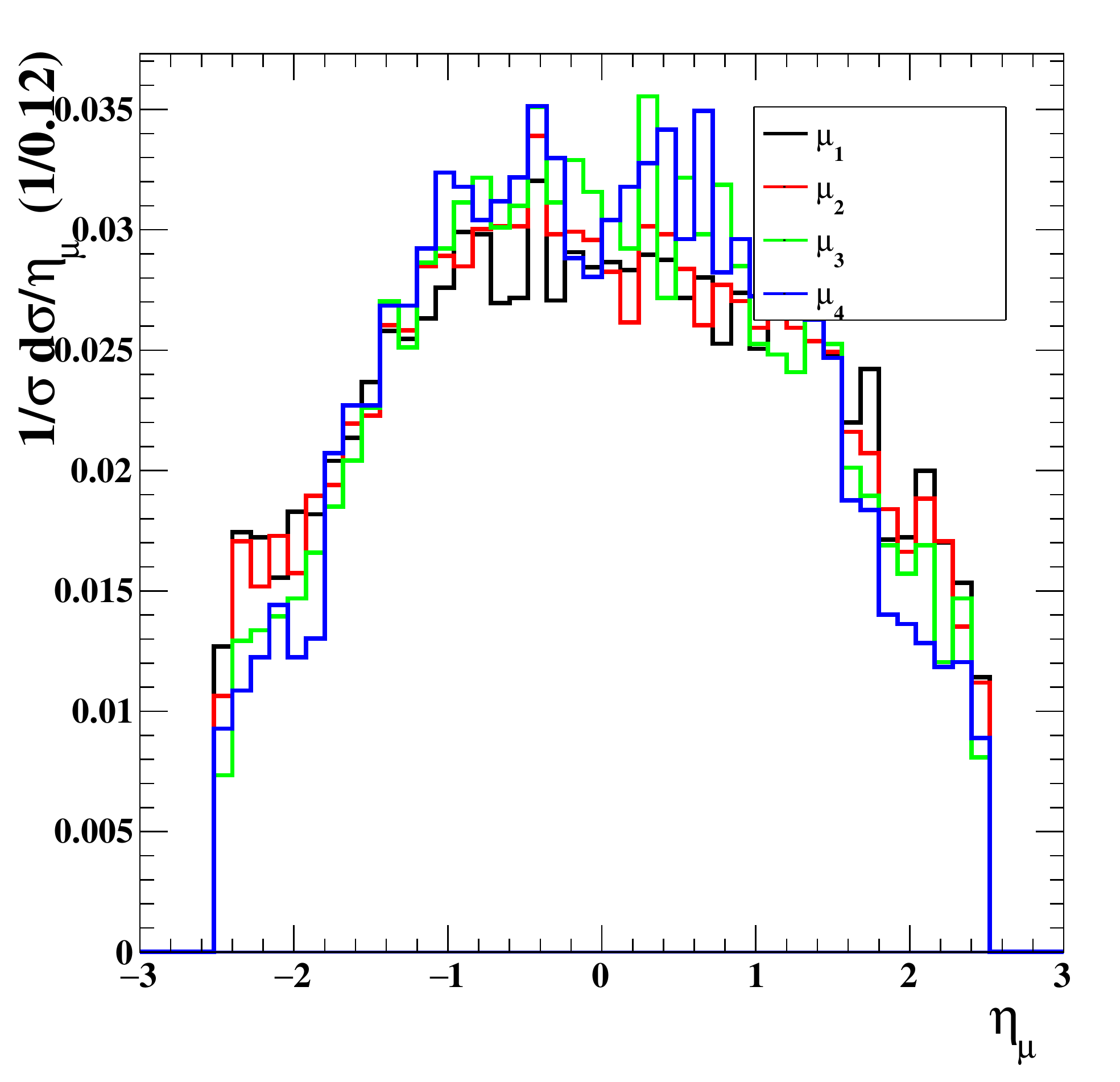}
\caption{\small \label{hs_pt_eta} 
Transverse momentum $p_{T_{\mu}}$ (left panel) and rapidity 
$\eta_\mu$ (right panel) distributions
for the four final state muons arranged in $p_T$ in the Higgs-portal model-1 
at LHC-14, $m_{h_s}$=0.5 GeV, at LHC 14 TeV with Delphes ATLAS simulations.
}
\end{figure}

\begin{figure}[th!]
\centering
\includegraphics[width=3in]{mj_m1_dR.pdf}\includegraphics[width=3in]{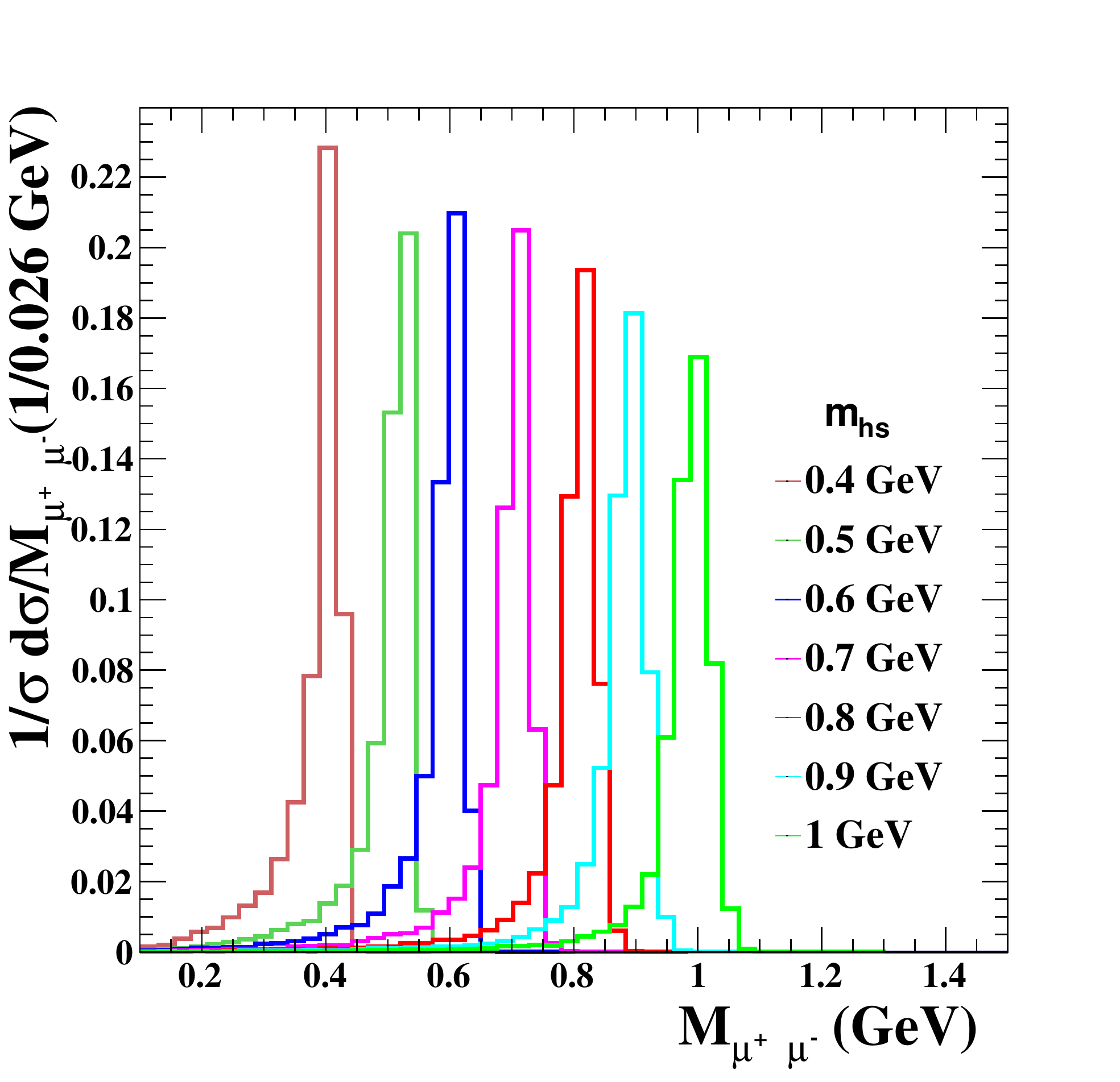}
\caption{\small \label{hs_im_dr} 
The opening angle $\Delta R_{\mu^+ \mu^-}$ (left panel) and 
the invariant mass distribution $M_{\mu^+ \mu^-}$ (right panel) for 
a pair of oppositely-charged muons inside a muon-jet for each benchmark 
point in the Higgs-portal model-1, at LHC 14 TeV with Delphes ATLAS
simulations.
}
\end{figure}

In the Higgs-portal model-1, there is only one light scalar boson in the dark
sector. 
We display the benchmark point $m_{h_s}$=0.5 GeV to show the $ p_{T} $ 
and $ \eta $ distributions in Fig.~\ref{hs_pt_eta} for the final state of two 
$2\mu$-jets, and the invariant mass distribution $M_{\mu^+ \mu^-}$ and 
the opening angle $\Delta R_{\mu^+ \mu^-}$ for a pair oppositely-charged 
muons inside a muon-jet in Fig.~\ref{hs_im_dr} for each benchmark point. 

\subsection{Higgs-portal model-2}

\begin{figure}[t!]
\centering
\includegraphics[width=3in]{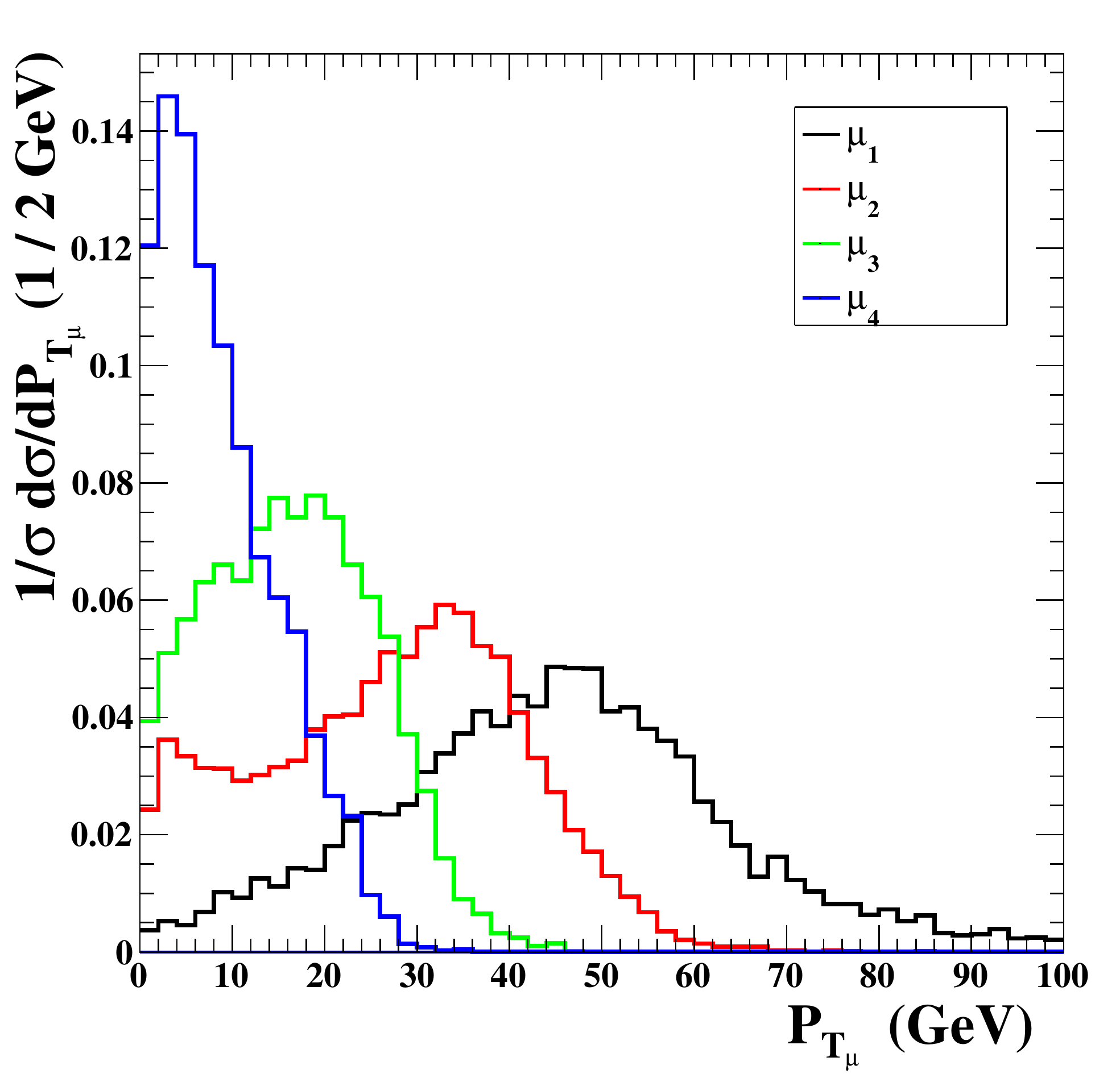}\includegraphics[width=3in]{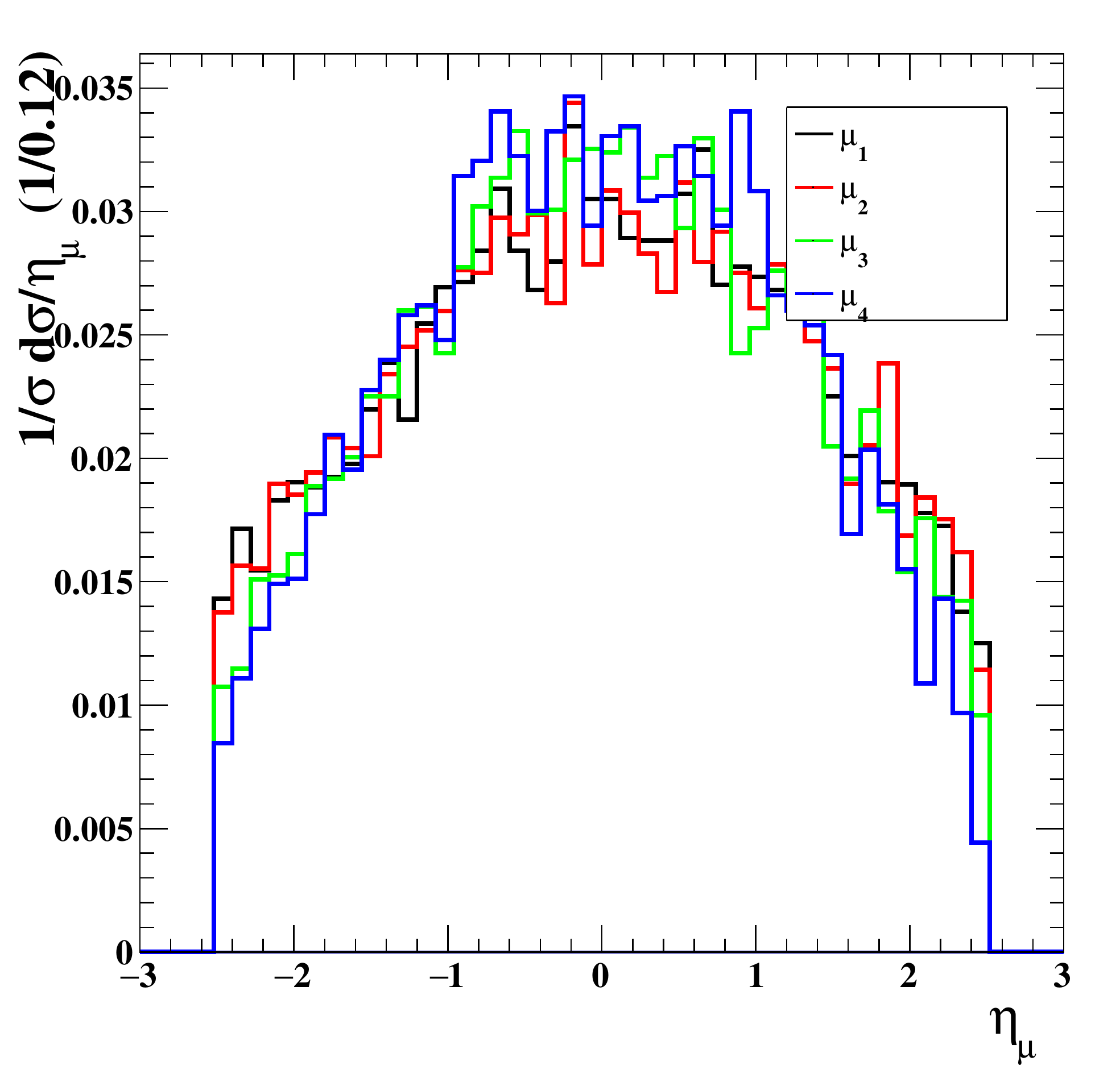}
\includegraphics[width=3in]{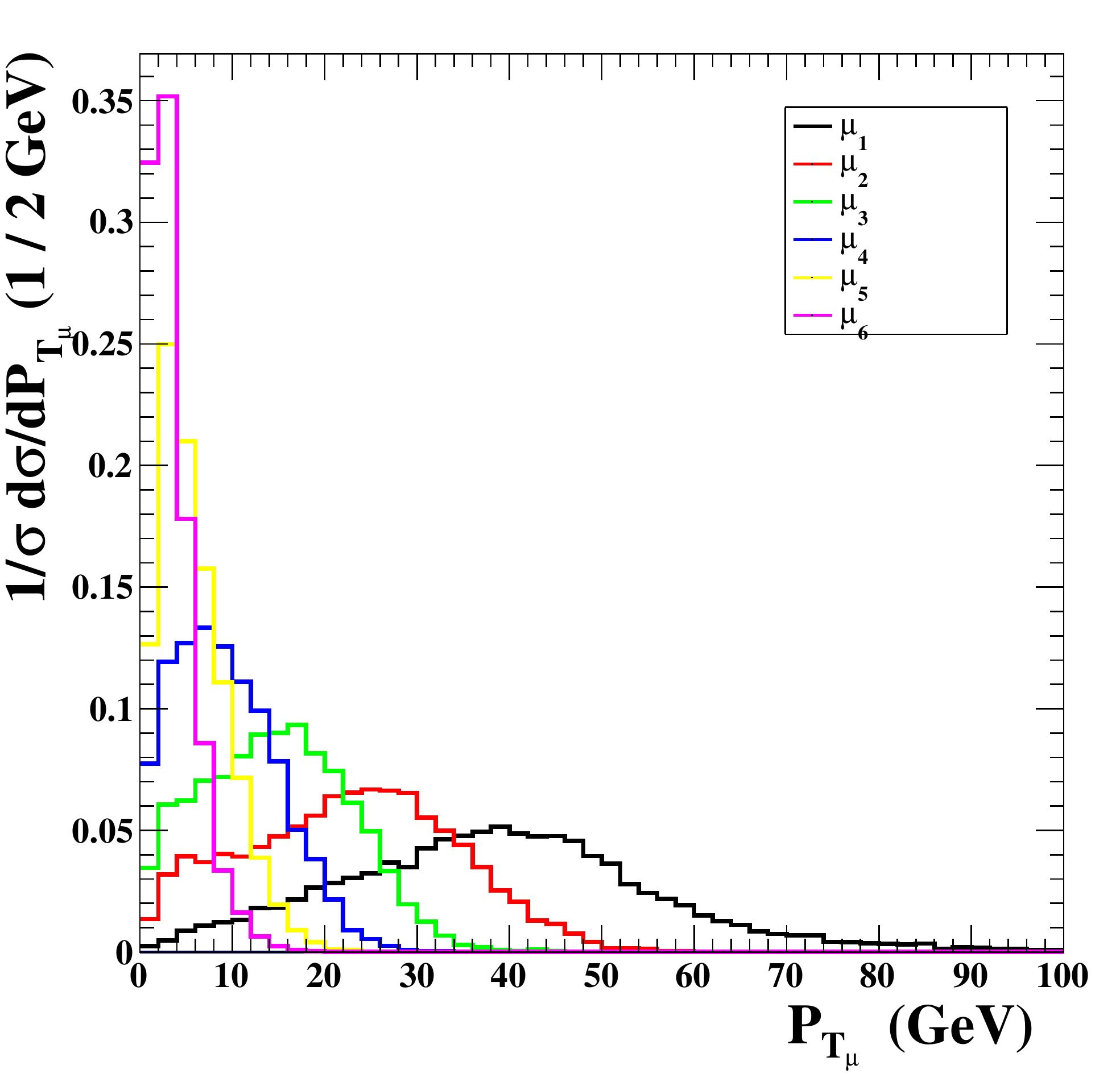}\includegraphics[width=3in]{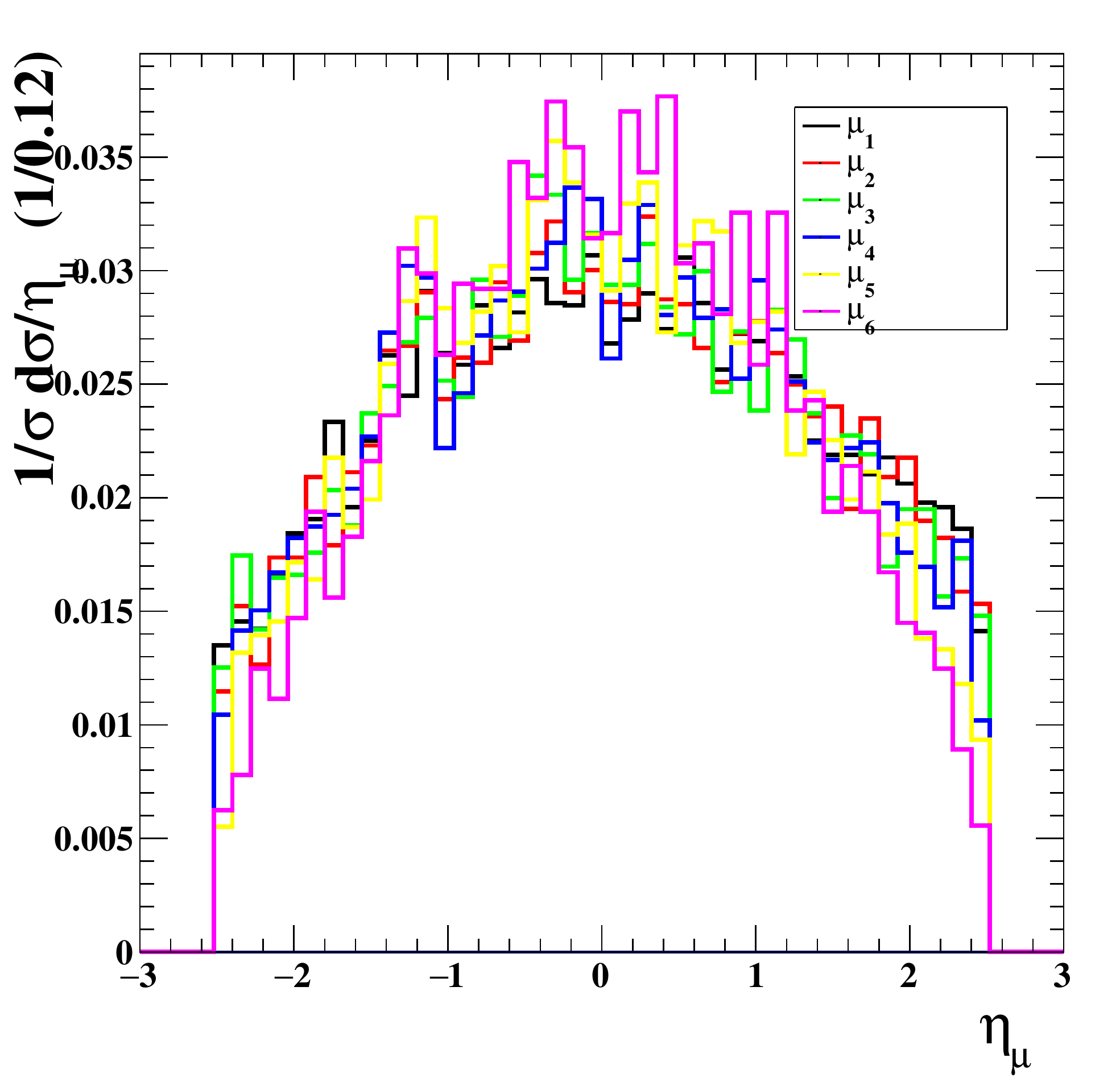}
\includegraphics[width=3in]{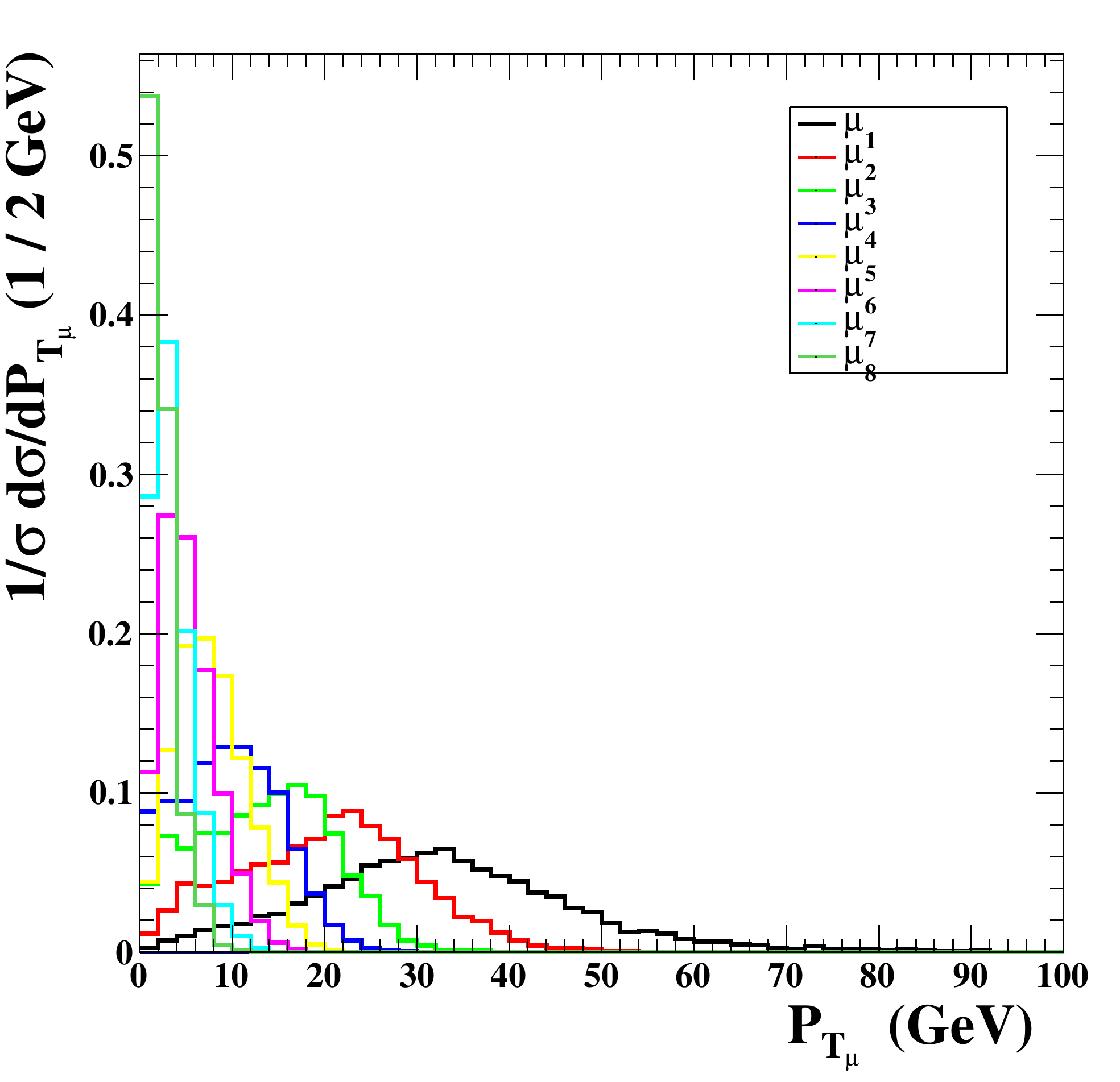}\includegraphics[width=3in]{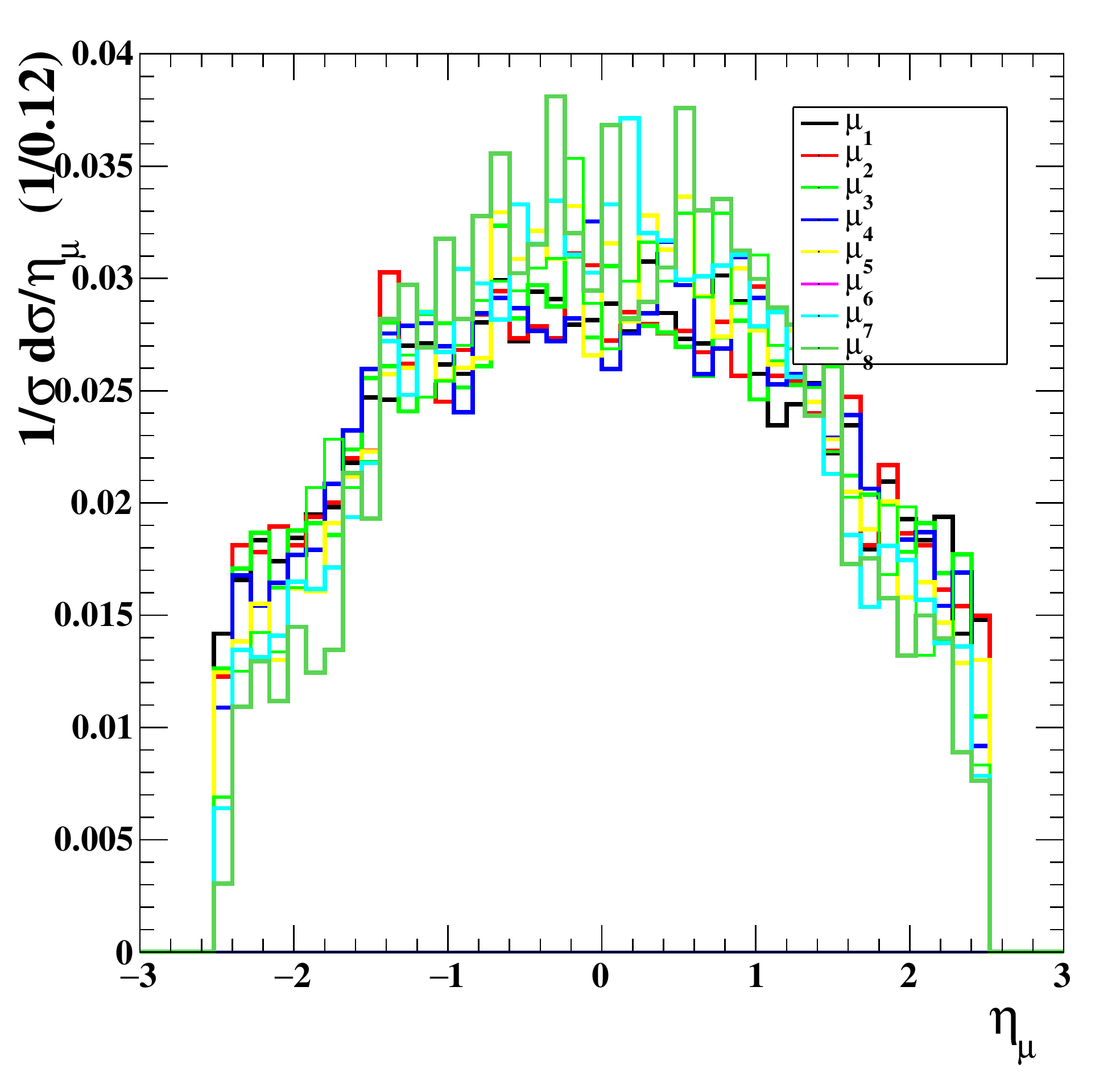}
\caption{\small \label{4m_6m_8m}
In the left panels: the $p_{T_\mu} $ distributions with the parameters of case 1, and $m_{h_{D_1}} $=2.5 GeV, $ m_{h_{D_2}} $= 0.5 GeV for final state 
topologies of TP1(upper), TP2(medium), and TP3(bottom) with muons
arranged in $p_T$. In the right panels : the corresponding $ \eta $ distributions. At LHC 14 TeV with Delphes ATLAS
simulations.
}
\end{figure}

\begin{figure}[h!]
\centering
\includegraphics[width=4.5in]{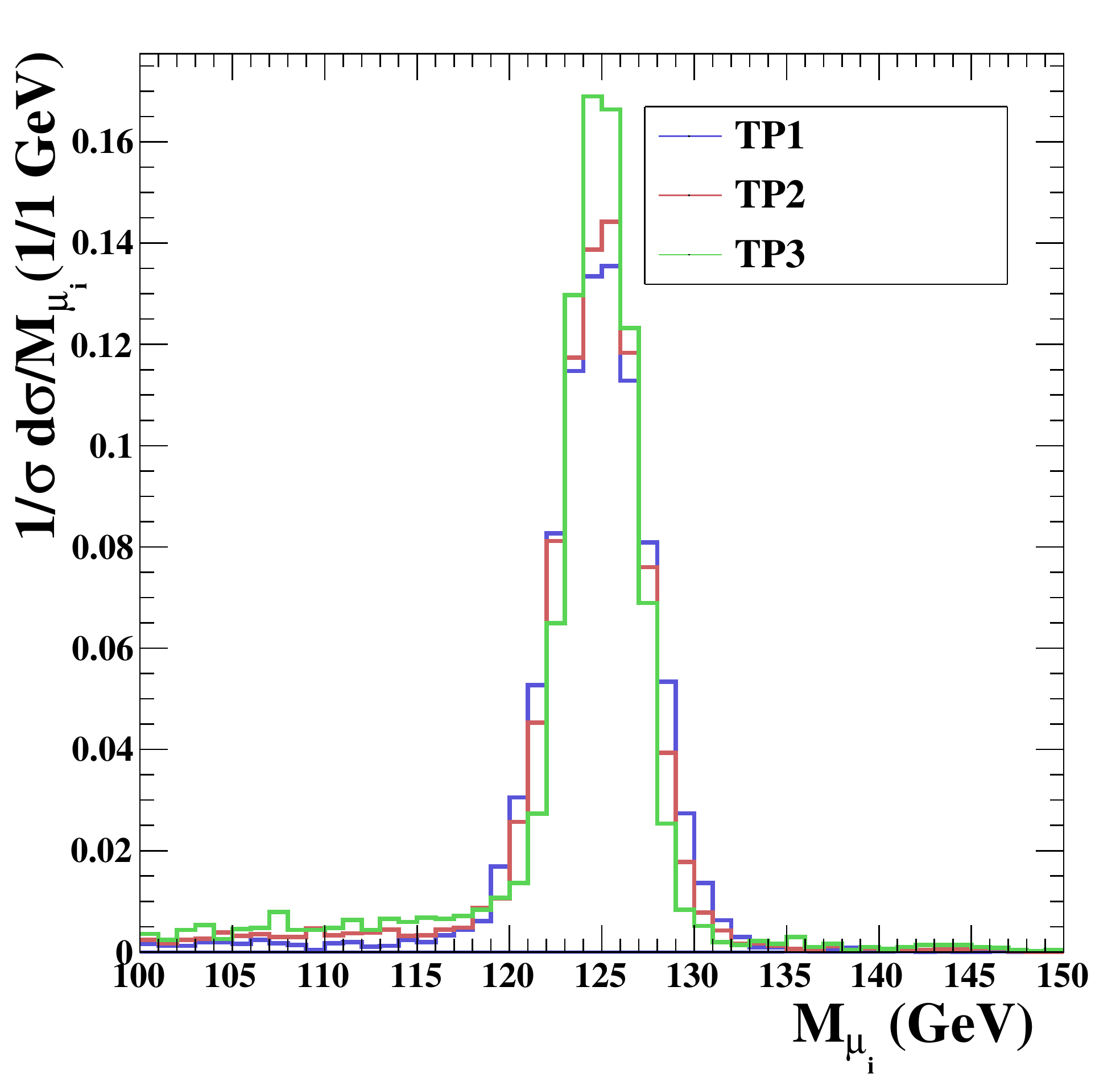}
\caption{\small \label{mh_eta}
The invariant mass distribution for the muon-jets, illustrated
for the Higgs-portal model-2, case 1:
$m_{h_{D_1}}=$2.5 GeV, $m_{h_{D_2}}=$1.0 GeV, at LHC 14 TeV with Delphes ATLAS
simulations. }
\end{figure}

We have explained the various event topologies in the current work and
they are
\begin{enumerate}
\item  TP1: two  $2\mu$-jets, 
\item  TP2: one $2\mu$-jet and one  $4\mu$-jet, and 
\item  TP3: two  $4\mu$-jets. 
\end{enumerate}
Note that the choice of parameters in case 1 allows all three event 
topologies. We show the $p_{T_\mu}$ and $\eta_\mu$ distributions 
for TP1, TP2, and TP3 using the case 1 parameters for the 
Higgs-portal model-2 in Fig.~\ref{4m_6m_8m}.  
In the Higgs-portal model, the light scalars comes
from the Higgs boson decay, thus the Higgs-mass-window cut can be used to
separate the signal from backgrounds. In Fig.~\ref{mh_eta}, we show the
invariant mass of $\mu-$jets for case 1 of the Higgs-portal model-2 to
illustrate the Higgs-mass window in three final-state topologies TP1,
TP2 and TP3.

\begin{figure}[th!]
\centering
\includegraphics[width=3in]{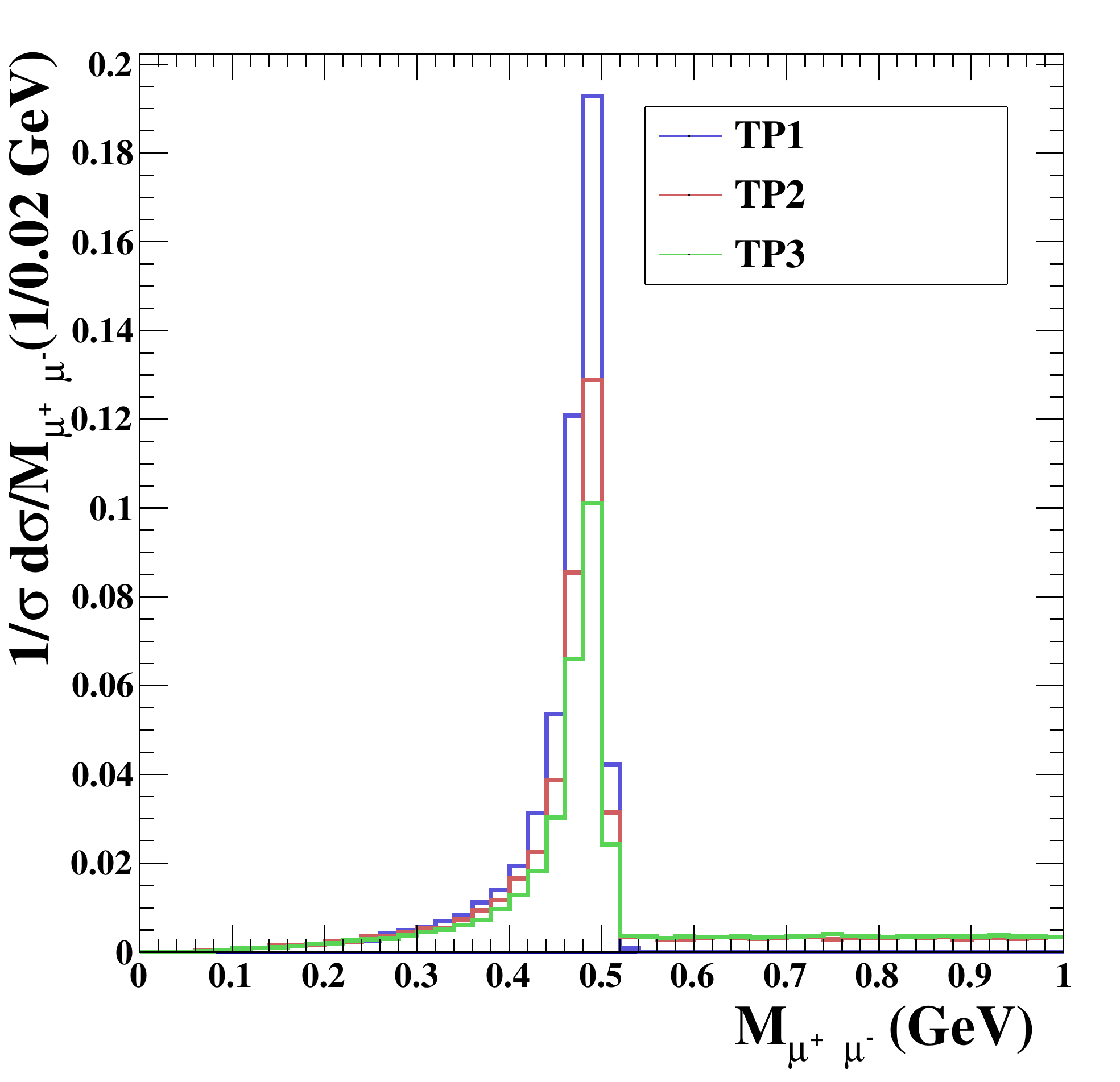}
\includegraphics[width=3in]{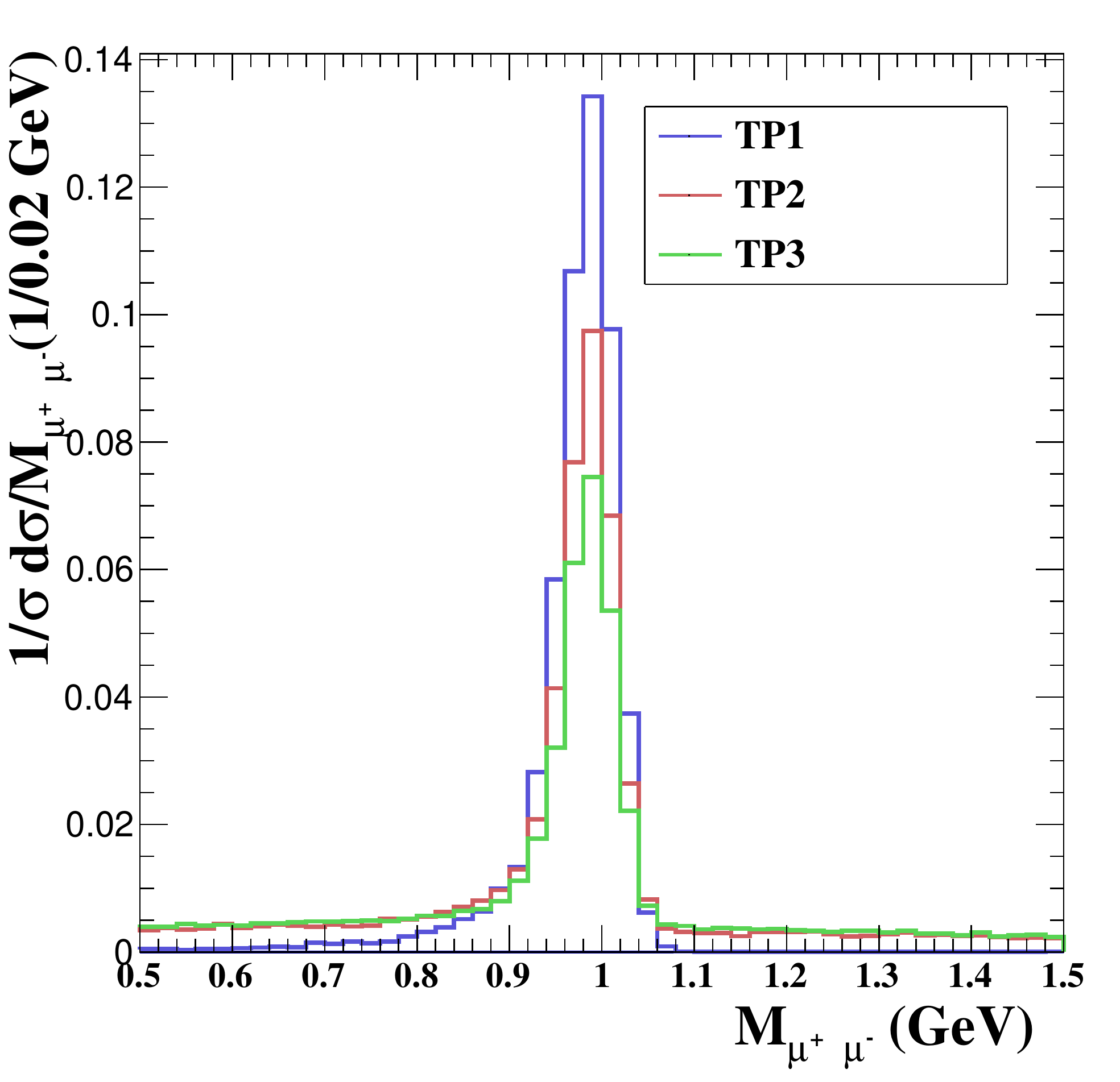}
\caption{\small \label{M_mu}
The invariant mass distribution $M_{\mu^+ \mu^-}$ for a pair of 
oppositely-charged  muons inside a muon-jet. We show the choice of 
parameters for
case 1 in the Higgs-portal model-2 with $m_{h_{D_1}} = 2.5$ GeV and 
$m_{h_{D_2}} = 0.5$ (left), 1.0 (right). At LHC 14 TeV with Delphes ATLAS simulations.
}
\end{figure}

In Fig.~\ref{M_mu}, we show the invariant mass distribution $M_{\mu^+ \mu^-}$ 
for a pair of oppositely-charged muons inside a muon-jet
in different final-state event topologies TP1, TP2, and TP3.
We show the choice of parameters for 
case 1 with $m_{h_{D_1}}=2.5$ GeV and $m_{h_{D_2}}=0.5/1.0$ GeV.
The invariant mass distributions in each case shown in Fig.~\ref{M_mu} 
clearly show the mass peaks of the light dark scalars for
different topologies.
For final-state topologies TP2 and TP3 
the $h_{D_1}$ will mostly
decay into $h_{D_2} h_{D_2}$, and so we can only see 
one mass peak at $m_{h_{D_2}}$ plus a long tail
because half of the times a wrong pair of oppositely-charged muons
are group together.

\section{Some detailed information about detectors}
The pixel detector of ATLAS or CMS is made up of a few layers of 
silicon pixels organized at radii of about a few cm to about 10 cm
\cite{pixel}.
The spatial resolution of the pixels ranges from $10-100\mu$m depending
on direction. Taking conservatively $100\mu$m as the spatial resolution
and divide it by the radius of the tracker, the angular resolution is 
of order $100\mu{\rm m}/10{\rm cm} \sim 10^{-3}$. This resolution is 
already better than the angle $0.01$ that we estimated above, so that 
the pixel detector can separate the very collimated muon-jet that we
consider in this work. 
However, there is no guarantee that the pattern recognition algorithms 
would be able to reconstruct two distinct tracks, 
especially in the presence of large number of pile-up events. 

Besides the inner pixel detector, the muon spectrometer is also
very important to identify and measure the momentum of muons.
The design of muon spectrometer in ATLAS and CMS is different.
The {\it Muon Spectrometer} of ATLAS is large in size but low
in magnetic field.  The advantages of this kind of design are its excellency
in stand-alone capabilities and safer for high multiplicities. 
Thus, the ATLAS muon detector 
performance is excellent over the whole $ \eta$ range and 
its resolution is nearly constant with $\eta $.
On the other hand, the CMS muon spectrometer is smaller
in size but high in magnetic field. The advantages of this kind of 
design are its superior combined momentum resolution in the central 
region and muons can be tracked and pointed back to the primary vertex. 
Therefore, the CMS muon performance
driven by the tracker is better near $ \eta\sim 0 $.
We specifically describe the ATLAS muon spectrometer in
the following. It is an extremely large tracking system,
consisting of three parts: 
(i) a magnetic field provided by three toroidal magnets,
(ii) a set of 1200 chambers measuring with high spatial precision 
the tracks of the outgoing muons, and 
(iii) a set of triggering chambers with accurate time-resolution. 
The extent of this sub-detector starts at a radius of 4.25 m close to
the calorimeters out to the full radius of the detector (11 m). Its
tremendous size is required to accurately measure the momentum of
muons, which first go through all the other elements of the detector
before reaching the muon spectrometer. It was designed to measure,
stand-alone, the momentum of 100 GeV muons with 3\% accuracy and of 1
TeV muons with 10\% accuracy. 
It also serves the function of simply identifying muons -- very few
particles of other types are expected to pass through the calorimeters
and subsequently leave signals in the Muon Spectrometer.

\section*{Acknowledgment}  
This work was supported the MoST of Taiwan under Grants 
No. NSC 102-2112-M-007-015-MY3 and MOST 105-2112-M-007-028-MY3.
S.C.H. is supported in parts by the National Science Foundation.

\end{document}